\documentclass[journal]{IEEEtran}
\ifCLASSINFOpdf
\else
\fi
\usepackage{cite}
\usepackage{amsmath}
\usepackage[pdftex]{graphicx} 
\usepackage{subfig}
\usepackage{array} 
\usepackage{fixltx2e}

\usepackage{booktabs} 
\usepackage{amssymb}
\usepackage{mathtools}
\usepackage{cancel} 
\usepackage{bm} 
\usepackage{esint} 
\usepackage{amsthm}
\usepackage{enumitem}

\usepackage{tikz}
\usepackage{tikz-3dplot}
\usetikzlibrary{backgrounds,3d,shadings,shapes.misc,decorations.pathmorphing,shapes,calc,fadings,shadows.blur}

\usepackage{amssymb,amsmath}

\usepackage{bm}
\newcommand{\bs}[1]{\bm{\mathrm{#1}}}

\newcommand{\vect}[1]{\vec{#1}}
\newcommand{\nhat}{\ensuremath{\hat{n}}}

\newcommand{\vecprime}[1]{\vect{#1}^{\,\prime}}

\renewcommand{\r}{\left(\vect{r}\right)}
\newcommand{\rp}{\left(\vecprime{r}\right)}

\newcommand{\rrpk}[1][]{\left(k_{#1}, \vect{r}, \vecprime{r}\right)}
\newcommand{\rrpki}[1][]{\left(k_i, \vect{r}, \vecprime{r}\right)}

\newcommand{\rvec}{\vect{r}}

\newcommand{\matr}[1]{\bs{#1}}

\newcommand{\abs}[1]{\left \lvert #1 \right\rvert }

\newcommand{\pvint}{\dashint}

\newcommand{\junk}[1] {}

\def\Xint#1{\mathchoice
{\XXint\displaystyle\textstyle{#1}}%
{\XXint\textstyle\scriptstyle{#1}}%
{\XXint\scriptstyle\scriptscriptstyle{#1}}%
{\XXint\scriptscriptstyle\scriptscriptstyle{#1}}%
\!\int}
\def\XXint#1#2#3{{\setbox0=\hbox{$#1{#2#3}{\int}$}
\vcenter{\hbox{$#2#3$}}\kern-.5\wd0}}

\def\dashint{\Xint-}

\newcommand*\widebar[1]{%
  \hbox{%
    \vbox{%
      \hrule height 0.5pt 
      \kern0.3ex
      \hbox{%
        \kern-0.05em
        \ensuremath{#1}%
        \kern-0.05em
      }%
    }%
  }%
} 

\renewcommand{\epsilon}{\varepsilon}

\newcommand{\opL}{\ensuremath{\mathcal{L}}} 
\newcommand{\opK}{\ensuremath{\mathcal{K}}} 

\newcommand{\outsub}{\ensuremath{0}}
\newcommand{\outsup}{\ensuremath{}}

\newcommand{\insub}{\ensuremath{i}}
\newcommand{\insup}{\ensuremath{}}

\newcommand{\insuball}{\ensuremath{}}
\newcommand{\insupall}{\ensuremath{}}

\newcommand{\Lout}[1][]{\ensuremath{\matr{L}_{\outsub#1}^{\outsup}}} 
\newcommand{\Kout}{\ensuremath{\matr{K}_{\outsub}^{\outsup}}} 
\newcommand{\Pxout}{\ensuremath{\matr{I}_{\times}}} 
\newcommand{\KPxout}{\ensuremath{\Kout + \frac{1}{2}\Pxout}} 

\newcommand{\Lmat}[1][]{\ensuremath{\matr{L}_{#1}}} 
\newcommand{\Kmat}[1][]{\ensuremath{\matr{K}_{#1}}} 

\newcommand{\Lin}{\ensuremath{\matr{L}_{\insub}^{\insup}}} 
\newcommand{\Kin}{\ensuremath{\matr{K}_{\insub}^{\insup}}} 
\newcommand{\Pxin}{\ensuremath{\matr{I}_{\times}}} 
\newcommand{\KPxin}{\ensuremath{\Kin - \frac{1}{2}\Pxin}} 

\newcommand{\Linall}{\ensuremath{\matr{L}_{\insuball}^{\insupall}}} 
\newcommand{\Kinall}{\ensuremath{\matr{K}_{\insuball}^{\insupall}}} 
\newcommand{\KPxinall}{\ensuremath{\Kinall - \frac{1}{2}\Pxin}} 

\newcommand{\LinNRP}{\ensuremath{\widetilde{\matr{L}}_{\insub}^{\insup}}} 
\newcommand{\KinNRP}{\ensuremath{\widetilde{\matr{K}}_{\insub}^{\insup}}} 
\newcommand{\LinNR}{\ensuremath{{\matr{L}}_{\insub,\mathrm{NR}}^{\insup}}} 
\newcommand{\KinNR}{\ensuremath{{\matr{K}}_{\insub,\mathrm{NR}}^{\insup}}} 
\newcommand{\KPxinNRP}{\ensuremath{\KinNRP - \frac{1}{2}\Pxin}} 

\newcommand{\figref}[1]{Fig.~\ref{#1}}
\newcommand{\secref}[1]{Section~\ref{#1}}

\usepackage{tcolorbox}

\usepackage{textpos}

\newcommand{\mySubtitle}[1]%
{%
	\begin{textblock}{14.0}(0.7, 2.9)
		\textbf{#1}%
	\end{textblock}%
}%

\newcommand{\redcol}{black}

\newcommand{\red}[1]{\textcolor{\redcol}{#1}}
\newcommand{\green}[1]{\textcolor{green!0!black}{#1}}

\newcommand{\Ecolor}[1]{{#1}}
\newcommand{\Hcolor}[1]{{#1}}

\newcommand{\Er}[1][]{\Ecolor{\vect{E}_{#1}\r}}

\newcommand{\Etr}[1][]{\Ecolor{\nhat_{#1} \times \vect{E}_{#1}\r}}
\newcommand{\Etrp}[1][]{\Ecolor{\nhat_{#1}' \times \vect{E}_{#1}\rp}}

\newcommand{\Emat}[1][]{\Ecolor{\matr{E}_{#1}}}

\newcommand{\Hr}[1][]{\Hcolor{\vect{H}_{#1}\r}}

\newcommand{\Htr}[1][]{\Hcolor{\nhat_{#1} \times \vect{H}_{#1}\r}}
\newcommand{\Htrp}[1][]{\Hcolor{\nhat_{#1}' \times \vect{H}_{#1}\rp}}
\newcommand{\Hmat}[1][]{\Hcolor{\matr{H}_{#1}}}

\newcommand{\Grrpk}[1][]{\ensuremath{\green{G_{#1}\rrpk}}}
\newcommand{\Grrpki}[1][]{\ensuremath{\green{G\rrpki}}}

\newcommand{\Grki}[1][]{\ensuremath{\green{G_{#1}\left(k_i, r\right)}}}

\newcommand{\rhomat}[1][]{\red{\matr{\rho}_{#1}}}

\makeatletter
\newlength\numerator@height
\newlength\frac@height
\newsavebox\numerator@box
\newsavebox\frac@box

\newcommand\dfracparens[3]{%
	\sbox{\numerator@box}{\ensuremath{#1}}%
	\sbox{\frac@box}{\ensuremath{\dfrac{#1}{#2}}}%
	\settoheight{\frac@height}{\usebox{\frac@box}}%
	\settoheight{\numerator@height}{\usebox{\numerator@box}}%
	\addtolength{\frac@height}{-\numerator@height}%
	\usebox{\frac@box}%
	\raisebox{\frac@height}{%
		\( \left( {#3} \right)
		\)}%
}
\makeatother

\begin{document}
%
\title{An Accelerated Surface Integral~Equation Method for the Electromagnetic~Modeling of Dielectric and Lossy Objects of Arbitrary Conductivity}
%
%
%

\author{Shashwat~Sharma,~\IEEEmembership{Student Member,~IEEE,}
        and~Piero~Triverio,~\IEEEmembership{Senior Member,~IEEE}
\thanks{Manuscript received $\ldots$; revised $\ldots$.}
\thanks{S. Sharma and P. Triverio are with the Edward S. Rogers Sr. Department of Electrical \& Computer Engineering, University of Toronto, Toronto,
ON, M5S 3G4 Canada, e-mails: shash.sharma@mail.utoronto.ca, piero.triverio@utoronto.ca.}
\thanks{This work was supported by the Natural Sciences and Engineering Research 
	Council of Canada (Collaborative Research and Development Grants 
	program), by Advanced Micro Devices, and by CMC Microsystems.}}

%
%

\markboth{IEEE Transactions on Antennas and Propagation}%
{Sharma \MakeLowercase{\textit{et al.}}: Bare Demo of IEEEtran.cls for IEEE Journals}
%



\maketitle

\begin{abstract}
  Surface integral equation (SIE) methods are of great interest for the numerical solution of Maxwell's equations in the presence of homogeneous objects. However, existing SIE algorithms have limitations, either in terms of scalability, frequency range, or material properties. We present a scalable SIE algorithm based on the generalized impedance boundary condition which can efficiently handle, in a unified manner, both dielectrics and conductors over a wide range of conductivity, size and frequency. We devise an efficient strategy for the iterative solution of the resulting equations, with efficient preconditioners and an object-specific use of the adaptive integral method. With a rigorous error analysis, we demonstrate that the adaptive integral method can be applied over a wide range of frequencies and conductivities. Several numerical examples, drawn from different applications, demonstrate the accuracy and efficiency of the proposed algorithm.
\end{abstract}

\begin{IEEEkeywords}
Electromagnetic modeling, surface integral equations, adaptive integral method, penetrable media.
\end{IEEEkeywords}

%
\IEEEpeerreviewmaketitle

\section{Introduction}

\IEEEPARstart{T}{he} efficient electromagnetic modeling of penetrable media is an important challenge in a wide variety of applications.
For example, the quantification of signal integrity phenomena in high-speed interconnects and of antenna losses requires the accurate modeling of the frequency-dependent variation of skin depth in lossy conductors.
Likewise, the sub-wavelength unit cells that comprise metasurfaces may involve both dielectric and conductive elements.
Conductor losses impact the electromagnetic response of the unit cells, and therefore must be modeled accurately~\cite{SRRmetaCond}.

Finite element methods~\cite{FEMJin} and volumetric integral equations~\cite{VolIE01}, though robust, require a volumetric discretization of the structure, which leads to a large number of unknowns.
This is of particular concern for lossy conductors at high frequency, where an extremely fine mesh is required to capture the shrinking skin depth.
Surface integral equation~(SIE) techniques, based on the boundary element method, are an effective alternative since they only require a surface mesh on the interfaces between objects~\cite{gibson}.
This leads to a smaller number of unknowns compared to volumetric methods.
Full-wave modeling of penetrable objects with SIEs requires solving an interior problem to capture the physics within objects, and an exterior problem to capture coupling between them.

Existing SIE methods for dielectrics include the Poggio Miller Chang Harrington Wu Tsai (PMCHWT) algorithm~\cite{PMCHWT02,PMCHWT03,PMCHWT04} and related formulations.
The methods in this class require adding together integral equations for adjacent media, which leads to loss of accuracy when there is a large contrast in electrical properties.
This renders PMCHWT-based techniques unsuitable for modeling lossy conductors embedded in a dielectric substrate.

For conductive media, approximate techniques such as the Leontovich surface impedance boundary condition (SIBC)~\cite{SIBC} are widely used, but are accurate only at high frequency.
The generalized impedance boundary condition (GIBC)~\cite{GIBC}, single-layer impedance matrix (SLIM)~\cite{AWPL2020} and the differential surface admittance~(DSA) approach~\cite{DSA01} have been proposed for modeling lossy conductors\cite{AWPL2020,DSA08,EPEPS2017} and dielectrics~\cite{UTK_AWPL2017}, but they require the inversion of at least one dense matrix for every object in the system.
This is only feasible for small objects or at high frequency, where skin effect has developed and the interior problem matrices are sparse.
Therefore, assembling the matrices which model the interior region becomes prohibitive, in terms of both CPU time and memory consumption, for large conductive objects at low frequency, and for dielectric objects in any regime.

The augmented electric field integral equation (AEFIE)~\cite{aefie2} avoids the low-frequency issues associated with the conventional electric field integral equation (EFIE)~\cite{lfbreakdown} by taking the charge density as an additional unknown.
Initially introduced for perfect electric conductors~\cite{aefie2}, the AEFIE was also extended to model penetrable media.
The enhanced augmented electric field integral equation (eAEFIE) formulation was proposed for both dielectrics~\cite{eaefie01} and conductors~\cite{eaefie02}.
The eAEFIE takes both electric and magnetic current densities as unknowns, in addition to the charge density, leading to a system of equations larger than those in GIBC- and DSA-based formulations.
The system matrix that results from the eAEFIE tends to be poorly conditioned, unless expensive dual basis functions are employed~\cite{BCorig,DualChew}.
Although the eAEFIE formulation avoids the inversion of matrices associated with the interior problem, the number of iterations required by an iterative solver is significantly increased when dual basis functions are not used, as we will show in \secref{sec:results}.
Furthermore, for conductive media, the eAEFIE only accounts for electromagnetic interactions between mesh elements that are close to each other~\cite{eaefie02}.
At low frequencies, where skin depth is large, ignoring far-away interactions compromises accuracy.

A SIE formulation for penetrable media was also proposed in~\cite{Song2003}, but is similar to the eAEFIE in that both electric and magnetic current densities are taken as unknown in the final system of equations.
A formulation similar to the GIBC was also proposed in~\cite{gibcHmatDanJiao}, which is based on~\cite{Song2003} and uses nested hierarchical matrices along with a dense matrix direct solver to accelerate both the interior and exterior problems.
This method achieves an overall time complexity of $\mathcal{O}(N)$, and has shown significant promise for the extraction of port parameters in on-chip applications such as interconnects and inductor coils.
However,~\cite{gibcHmatDanJiao} considers only conductive objects, and might not be suitable for problems such as dielectric antennas and metasurfaces.

In summary, existing SIE techniques tend to be effective only under restrictive assumptions on material properties and/or frequency, or require advanced basis functions which lead to an increased computational cost.
Their limitations may be an issue in practical applications, since many devices of great interest involve both dielectrics and conductors, and must be analyzed over a wide range of frequencies or, equivalently, involve objects of different electrical size.

In this paper, we propose a SIE technique which is based on the GIBC formulation~\cite{GIBC}, but reduces the computational complexity of the interior problem from $\mathcal{O}(N^3)$ to approximately $\mathcal{O}(N^{1.5}\log N)$, without requiring dual basis functions.
The proposed method handles, in a unified manner, both dielectrics and lossy conductors over a wide range of conductivity, size and frequency.
A numerical surface impedance operator represented in compressed form is devised, which accurately models the interior physics of each object in the structure, but avoids the assembly and factorization of large dense matrices, unlike the original GIBC.
To accelerate matrix inversions related to the impedance operator, we develop a multiple-grid adaptive integral method (AIM)~\cite{AIMbles} for both dielectric and conductive media, by constructing multiple object-specific AIM grids.
Matrix-vector products involving the impedance operator are efficiently computed with an iterative solver, and further accelerated with a preconditioner based on the interactions of nearby mesh elements.
The proposed operator and preconditioner naturally take advantage of matrix sparsity arising from skin effect, when present.
Unlike the eAEFIE formulation~\cite{eaefie01}, the proposed method does not require expensive dual basis functions to achieve good conditioning, nor does it require taking both electric and magnetic current densities as unknowns.

The second contribution of this paper is a systematic error analysis of the AIM in the context of conductive media.
It is demonstrated that the AIM can be applied to lossy media across wide ranges of frequency and conductivity, with a grid resolution based only on frequency and the real part of the object's permittivity.
In particular, it is shown that the AIM can be used for highly conductive media over a wide range of frequency, without the need for an extremely fine mesh.
The proposed method is therefore suitable for applications which require accurate modeling of the frequency-dependent skin effect in conductors.
An analysis of the AIM for highly lossy media has not been presented before, to the best of our knowledge.
  
The third contribution of this manuscript is a detailed comparison of the conditioning and performance of the GIBC, the eAEFIE, and the proposed method.
A direct comparison of matrix conditioning between state-of-the-art penetrable medium formulations has not been offered previously.
We demonstrate that the proposed technique leads to faster convergence of iterative solvers than the eAEFIE~\cite{eaefie01, eaefie02}, when dual basis functions are not used.

This paper is organized as follows. \secref{sec:setup} defines the notation and general setup which will be used throughout.
\secref{sec:proposed} describes the proposed algorithm.
The accuracy of AIM for general lossy media is established in \secref{sec:objAIM}.
Finally, numerical results demonstrating the efficiency of the proposed method are presented in \secref{sec:results}, and concluding remarks are provided in \secref{sec:concl}.


\section{Setup and Notation}\label{sec:setup}

Throughout this work, we consider time-harmonic fields. A time dependence of $e^{j\omega t}$ is assumed and suppressed. Field quantities are written with an overhead arrow, for example
$\vect{a}\r$. Primed coordinates represent source points, while unprimed coordinates represent observation points. Matrices
and column vectors are written in bold letters, such as $\matr{A}$.

\begin{figure}[t]
  \centering
  \begin{tikzpicture}[scale = 1, >=latex]
	\draw[black, fill = black!10] (0,0) ellipse (3.0cm and 1.3cm);
	\draw[black, dashed, fill = none] (0,0) ellipse (2.9cm and 1.2cm);
	\draw[black, dashed, fill = none] (0,0) ellipse (3.1cm and 1.4cm);
	\node at (3.0, 0.2) [right] {$\mathcal{S}_i^+$};
	\node at (2.9, 0.2) [left] {$\mathcal{S}_i^-$};
	\draw[black, ->, line width = 0.4mm] (0, -1.3) -- (0, -1.9);
	\node at (0, -1.9) [right] {${\nhat}_i$};
	\draw[black, ->, line width = 0.4mm] (0, 1.3) -- (0, 0.7);
	\node at (0, 0.7) [right] {${\nhat}_0$};
	\node at (0.0,-0.5) {$({\mu_i}, {\epsilon_i}, \sigma_i)$};
	\node at (0.0,0.4) [below] {$\vect{E}_i(\vect{r}), \vect{H}_i(\vect{r})$};
	\node at (-1.4,0.8) [left]{${\mathcal{V}_i}$};
	\draw[black, thick] (-3.0cm,0.0cm)--+(0.4,0);
	\node at (-2.0,0.0) [left]{${\mathcal{S}_i}$};
	\node at (-3.0,0.9) [left]{${\mathcal{V}_0}$};
	\node at (-2.6,-1.4) [below] {$\vect{E}_0(\vect{r}), \vect{H}_0(\vect{r})$};
	\node at (2.2, -1.7) {$(\mu_0,\epsilon_0)$};
	\node at (-2.6,1.9) [below] {$\vect{E}_\mathrm{inc}(\vect{r}), \vect{H}_\mathrm{inc}(\vect{r})$};
	%
\end{tikzpicture}
  \caption{Geometric setup considered in this work.}\label{fig:geom}
\end{figure}
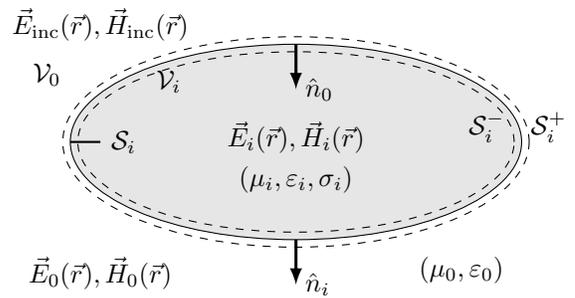

We consider a structure composed of $N_\mathrm{obj}$ objects, each occupying volume $\mathcal{V}_i$, $i = 1,2\ldots N_\mathrm{obj}$, and $\mathcal{V}~\triangleq~\bigcup_{i=1}^{N_\mathrm{obj}} \mathcal{V}_i$.
Each object is bounded by a surface $\mathcal{S}_i$, and we define $\mathcal{S}~\triangleq~\bigcup_{i=1}^{N_\mathrm{obj}} \mathcal{S}_i$.
Each surface $\mathcal{S}_i$ has outward unit normal vector $\nhat_i$ pointing out of $\mathcal{V}_i$.
Points infinitesimally close to the surface $\mathcal{S}$ outside $\mathcal{V}$ are denoted as $\mathcal{S}^+$; those slightly inside $\mathcal{V}$ are denoted as $\mathcal{S}^-$.
The objects are embedded in a homogeneous background medium denoted by index $i = 0$, so that the surface $\mathcal{S}_0$ coincides with $\mathcal{S}$ but with its unit normal vector $\nhat_0$ pointing into $\mathcal{V}$.
The setup described above is visualized in \figref{fig:geom}.

Each object in the structure is assumed to be homogeneous with permittivity $\epsilon_i$, permeability $\mu_i$, and electrical conductivity $\sigma_i$.
The complex wave number and wave impedance of the $i^\mathrm{th}$ object are
\begin{align}
  k_i = \sqrt{\omega\mu_i \left(\omega\epsilon_i - j\sigma_i\right)} \label{eq:kin}
\end{align}
and
\begin{align}
  \eta_i = \sqrt{\dfrac{\omega\mu_i}{\left(\omega\epsilon_i - j\sigma_i\right)}}\label{eq:etain}
\end{align}
respectively, where $\omega$ is the cyclical frequency, and ${j = \sqrt{-1}}$.
The system may be excited with an incident plane wave, ${[\Er[\mathrm{inc}], \Hr[\mathrm{inc}]]}$, ${\vect{r} \in \mathcal{V}_0}$.
This leads to the field distributions ${[\Er[i], \Hr[i]]}$, ${\vect{r} \in \mathcal{V}_i}$, for ${i = 0,1\ldots N_\mathrm{obj}}$.
A port excitation may also be used in the proposed method, such as the one discussed in~\cite{gope,CPMT2019arxiv} and used in this manuscript, or the one proposed in~\cite{Song2003,gibcHmatDanJiao}.

\section{Proposed Method}\label{sec:proposed}


\subsection{Interior Problem}\label{sec:interior}

Following the GIBC formulation~\cite{GIBC}, the interior problem for the $i^\mathrm{th}$ object in the structure is described via the magnetic field integral equation~(MFIE)~\cite{gibson},
\begin{multline}
  -j\omega\epsilon_i\, \nhat_i \times \opL_i \left[\Etrp[i]\right]\r \\
  - \nhat_i \times \opK_i \left[\Htrp[i]\right]\r
  + \dfrac{1}{2} \Htr[i] = 0 \label{eq:MFIE1}
\end{multline}
for $\vect{r} \in \mathcal{S}_i^+,\, \vect{r}\,' \in \mathcal{S}_i^-$, and $i = 1,2\ldots N_\mathrm{obj}$.
The integro-differential operators $\opL_i$ and $\opK_i$ are defined as~\cite{gibson}
\begin{multline}
  \opL_i\left[\vect{X}\rp\right]\r \\= \int_\mathcal{S} \left( \vect{X}\rp + \dfrac{1}{k_i^2}\nabla'\nabla' \cdot \vect{X}\rp \right) \Grrpki[i]\, d\mathcal{S},\label{eq:opL}
\end{multline}
\begin{align}
  \opK_i\left[\vect{X}\rp\right]\r = \pvint_\mathcal{S} \nabla\Grrpki[i] \times \vect{X}\rp\, d\mathcal{S}, \label{eq:opK}
\end{align}
where the homogeneous Green's function of medium $i$ is
\begin{align}
  \Grrpki[i] = \dfrac{e^{-j k_i r}}{4\pi r}, \label{eq:Grrpki}
\end{align}
where $r = |\vect{r} - \vect{r}^{\,'}|$.
The integral in~\eqref{eq:opK} is evaluated in the principal value sense~\cite{gibson}.

A triangular mesh is generated for the surface of each object.
Equation~\eqref{eq:MFIE1} is then discretized by expanding $\Etr[i]$ and $\Htr[i]$ with $\mathrm{RWG}$ basis functions~\cite{RWG} and tested with rotated $\nhat_i \times \mathrm{RWG}$ functions.
This leads to the discretized equation
\begin{align}
  -j\omega\epsilon_i\, \Lin \Emat[i] - \left( \KPxin \right) \Hmat[i] &= \matr{0}, \label{eq:MFIE1dis}
\end{align}
where $\matr{L}$ and $\matr{K}$ are the discretized $\opL$ and $\opK$ operators whose subscripts denote the medium whose Green's function is used.
Matrix $\Pxout$ is the identity operator obtained by testing $\mathrm{RWG}$ functions with rotated $\mathrm{RWG}$ functions.
Column vectors $\Emat$ and $\Hmat$ read
\begin{align}
  \Emat &= \begin{bmatrix} {\Emat[1]}^T & {\Emat[2]}^T & \cdots & {\Emat[i]}^T & \cdots & {\Emat[N_\mathrm{obj}]}^T \end{bmatrix}^T, \label{eq:unkE}\\
  \Hmat &= \begin{bmatrix} {\Hmat[1]}^T & {\Hmat[2]}^T & \cdots & {\Hmat[i]}^T & \cdots & {\Hmat[N_\mathrm{obj}]}^T \end{bmatrix}^T, \label{eq:unkH}
\end{align}
where $\Emat[i]$ and $\Hmat[i]$ contain the coefficients of the basis functions associated with $\Etr[i]$ and $\Htr[i]$, respectively.

\subsection{Exterior Problem}\label{sec:exterior}

The exterior problem, responsible for capturing the coupling between objects, is described with the electric field integral equation (EFIE)~\cite{gibson},
\begin{multline}
  j\omega\mu_0\, \nhat_i \times \opL_0 \left[\Htrp[i]\right]\r
  - \nhat_i \times \opK_0 \left[\Etrp[i]\right]\r\\
  - \dfrac{1}{2} \Etr[i] = -\nhat_i \times \Er[\mathrm{inc}], \label{eq:EFIE01}
\end{multline}
where $\vect{r},\, \vect{r}\,' \in \mathcal{S}^+$, and $i = 1,2\ldots N_\mathrm{obj}$.
Expanding $\Etr[i]$ and $\Htr[i]$ with $\mathrm{RWG}$ basis functions, and testing with  $\nhat_i \times \mathrm{RWG}$ functions as before yields
\begin{align}
  j\omega\mu_0\, \Lout \Hmat - \left( \KPxout \right) \Emat = -\Emat[\mathrm{inc}], \label{eq:EFIE01dis}
\end{align}
where column vector $\Emat[\mathrm{inc}]$ relates to the incident electric field $\nhat_i \times \Er[\mathrm{inc}]$.
The choice of using the MFIE in the interior problem and the EFIE in the exterior problem leads to a well conditioned final system, as demonstrated later in \secref{sec:results}.

\subsection{Existing Methods and Gaps}\label{sec:existing}

From~\eqref{eq:MFIE1dis} and~\eqref{eq:EFIE01dis}, two state-of-the-art formulations for modeling penetrable media can be derived: the GIBC~\cite{GIBC} and the eAEFIE~\cite{eaefie01,eaefie02}.
We will briefly review their strengths and limitations, to motivate the choice of the proposed formulation.

\subsubsection{GIBC}\label{sec:existingGIBC}
The electric field coefficient vector $\Emat[i]$ in~\eqref{eq:MFIE1dis} can be solved for with the rearrangement
\begin{align}
  \Emat[i] = \dfrac{-1}{j\omega\epsilon_i} {\left( \Lin \right)}^{-1} \left( \KPxin \right) \Hmat[i] = \matr{Z}_{i} \Hmat[i], \label{eq:GIBCdef}
\end{align}
where it is assumed $\omega \neq 0$, and $i = 1,2\ldots N_\mathrm{obj}$.
The resulting matrix $\matr{Z}_{i}$ is the GIBC operator for $\mathcal{V}_i$~\cite{GIBC}.
The GIBC matrix for all objects, $\matr{Z}$, can be assembled as
\begin{align}
  \matr{Z} = \mathrm{diag}\begin{bmatrix} \matr{Z}_{1} & \matr{Z}_{2} & \cdots & \matr{Z}_{i} & \cdots & \matr{Z}_{N_\mathrm{obj}} \end{bmatrix}. \label{eq:glGIBCdef}
\end{align}
Using~\eqref{eq:GIBCdef} in~\eqref{eq:EFIE01dis} allows us to eliminate $\Emat$ to get the final system of equations,
\begin{align}
  \left[ j\omega\mu_0\, \Lout - \left( \KPxout \right) \matr{Z} \right] \Hmat = -\Emat[\mathrm{inc}]. \label{eq:GIBCsys}
\end{align}
The difficulty in devising a scalable GIBC formulation stems from the complexity of the system matrix in~\eqref{eq:GIBCsys}, which includes the inverse of the $\Lin$ matrices in~\eqref{eq:GIBCdef}.
This complexity makes the application of iterative solvers, preconditioners, and FFT-based acceleration schemes such as the AIM~\cite{AIMbles} quite challenging.
Therefore, the GIBC is feasible only for small objects, or for conductors at high frequencies where skin effect makes $\Lin$ and $\Kin$ sparse.
The GIBC becomes prohibitively expensive for large dielectrics, and for large conductive objects at low and intermediate frequencies, where far-away interactions cannot be ignored.

\subsubsection{eAEFIE}\label{sec:existingeAEFIE}
In the eAEFIE, both $\Emat$ and $\Hmat$ are taken as unknowns to avoid the inversion of $\Lin$ in~\eqref{eq:GIBCdef}.
In the eAEFIE~\cite{eaefie01} and in the numerical examples in \secref{sec:results}, the AEFIE is used to describe both interior and exterior problems.
However, for simplicity we will consider the case of the standard EFIE in this section,
\begin{align}
  \begin{bmatrix}
    j\omega\mu_0\, \Lout & -\left( \KPxout \right) \\
    j\omega\, \Linall &	-\left( \KPxinall \right)
  \end{bmatrix}
                        \begin{bmatrix}
                          \Hmat \\ \Emat
                        \end{bmatrix} =
  \begin{bmatrix}
    -\Emat[\mathrm{inc}] \\ \matr{0}
  \end{bmatrix} \label{eq:eAEFIEsysEE}
\end{align}
where
\begin{multline}
  \Linall =\, 
  \mathrm{diag}\begin{bmatrix}
    \mu_1\matr{I} & \mu_2\matr{I} & \cdots & \mu_{N_\mathrm{obj}}\matr{I}
  \end{bmatrix}\cdot\\
  \mathrm{diag}\begin{bmatrix}
    \matr{L}_1 & \matr{L}_2 & \cdots & \matr{L}_{N_\mathrm{obj}}
  \end{bmatrix}, \label{eq:LinallE}
\end{multline}
and
\begin{align}
  \Kinall = \mathrm{diag}\begin{bmatrix}
    \matr{K}_1 & \matr{K}_2 & \cdots & \matr{K}_{N_\mathrm{obj}}
  \end{bmatrix}. \label{eq:Kinall}
\end{align}
When a conventional discretization with $\mathrm{RWG}$ and rotated $\mathrm{RWG}$ functions is used, the system matrix in~\eqref{eq:eAEFIEsysEE} is poorly conditioned compared to~\eqref{eq:GIBCsys} due to the rank deficiency of $\Pxin$~\cite{BCcalderon}.
Therefore, dual basis functions were used to expand $\Emat$ in~\cite{eaefie01,BCorig} to improve the conditioning of~\eqref{eq:eAEFIEsysEE}.
A barycentric dual mesh is generated such that $6$ small dual triangles are inscribed within each ``parent'' primary triangle.
While dual basis functions improve conditioning, leading to a significant reduction in the number of iterations required for solving~\eqref{eq:eAEFIEsysEE}, the cost of generating the $\Kout$ and $\Kmat$ matrices may increase by up to a factor of $6\times$ due to the dual mesh.
Introducing the dual mesh and basis functions also significantly increases code complexity, and the refined nature of the dual mesh may increase the number of triangles with poor aspect ratio for complicated or multiscale structures.
For these reasons, it is desirable to explore formulations which are efficient and well conditioned even without employing dual basis functions.

The conditioning of~\eqref{eq:eAEFIEsysEE} with a conventional discretization can be improved by using the MFIE in the interior problem, 
\begin{align}
  \begin{bmatrix}
    j\omega\mu_0\, \Lout & - \left( \KPxout \right) \\
    - \left( \KPxinall \right) & -j\omega\, \Linall
  \end{bmatrix}
                                 \begin{bmatrix}
                                   \Hmat \\ \Emat
                                 \end{bmatrix} =
  \begin{bmatrix}
    -\Emat[\mathrm{inc}] \\ \matr{0}
  \end{bmatrix}, \label{eq:eAEFIEsys}
\end{align}
where the well-conditioned $\Lin$ matrices are now along the diagonal.
In this case, $\Linall$ is defined as
\begin{multline}
  \Linall = 
  \mathrm{diag}\begin{bmatrix}
    \epsilon_1\matr{I} & \epsilon_2\matr{I} & \cdots & \epsilon_{N_\mathrm{obj}}\matr{I}
  \end{bmatrix}\cdot\\
  \mathrm{diag}\begin{bmatrix}
    \matr{L}_1 & \matr{L}_2 & \cdots & \matr{L}_{N_\mathrm{obj}}
  \end{bmatrix}. \label{eq:Linall}
\end{multline}
While the simple structure of the system matrix of~\eqref{eq:eAEFIEsys} makes the application of iterative solvers and FFT-based acceleration techniques possible~\cite{eaefie01}, the conditioning of~\eqref{eq:eAEFIEsys} is still inferior to that of~\eqref{eq:GIBCsys}, as will be demonstrated in \secref{sec:results}.

Thus, both the GIBC and eAEFIE have limitations, either in performance or matrix conditioning.

\subsection{Proposed Formulation}\label{sec:implicit}

We devise an accelerated GIBC algorithm where the interior problem is modeled in $\mathcal{O}(N^{1.5}\log N)$ time.
The proposed SIE method is applicable to both dielectric and conductive objects of any size (unlike the original GIBC) and enjoys good conditioning over a wide frequency range without the use of expensive dual basis functions (unlike the eAEFIE). 

Our starting point is the GIBC formulation~\eqref{eq:GIBCdef}, because of its superior conditioning.
The scalability limitations of the original scheme arise from two issues:
\begin{enumerate}
\item For conductors at low frequency, and for dielectric objects at any frequency, the matrices $\Lin$, $\Kin$ and $\matr{Z}_i$ are dense, and become prohibitively expensive to compute and store as object size increases.
\item If an iterative solver is used to both solve~\eqref{eq:GIBCsys} and to apply ${(\Lin)}^{-1}$ in~\eqref{eq:GIBCdef}, the second iterative process will be nested into the first one, potentially leading to high computational cost unless very efficient preconditioners and matrix-vector product algorithms are devised for both iterative solutions.
\end{enumerate}
The following subsections describe how each of these limitations is alleviated with the proposed method.

\subsubsection{AIM Acceleration for the Interior Problem Matrices}

To overcome the first issue, we demonstrate that the AIM can be successfully used to compute matrix-vector products involving matrices $\Lin$, $\Kin$ and $\matr{Z}_i$ in a scalable way.
These matrices describe electromagnetic effects inside each object.
Since objects can range from a perfect dielectric to a highly-conductive body, and have very different geometrical scales, we use independent Cartesian grids to apply the AIM to each object.
In this way, the resolution of each grid can be tailored to the geometry, wavelength, and skin depth in each object.
Furthermore, this approach minimizes the size of the AIM grid and therefore the cost of the associated operations (e.g. FFTs) required for each object. 

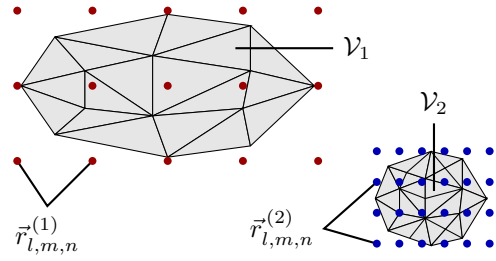
\begin{figure}[t]
  \centering
  \begin{tikzpicture}[scale = 1, >=latex]%
	\coordinate (mup1) at (-1.95,0.0);
	\coordinate (mup2) at (-1.1,0.2);
	\coordinate (mup3) at (-0.2,0.4);
	\coordinate (mup4) at (1.1,0.2);
	\coordinate (mup5) at (1.95,0.0);
	\coordinate (up1) at (-1.5,0.65);
	\coordinate (up2) at (0.0,0.95);
	\coordinate (up3) at (1.1,0.8);
	\coordinate (mlp2) at (-1.1,-0.3);
	\coordinate (mlp3) at (-0.2,-0.4);
	\coordinate (mlp4) at (1.0,-0.3);
	\coordinate (lp1) at (-1.5,-0.65);
	\coordinate (lp2) at (0.0,-0.95);
	\coordinate (lp3) at (1.1,-0.8);
	\coordinate (c1) at ($ 0.333*(mup1) + 0.333*(mup2) + 0.334*(up1) $);
	\coordinate (c2) at ($ 0.333*(mup3) + 0.333*(mup2) + 0.334*(up1) $);
	\coordinate (c3) at ($ 0.333*(mup3) + 0.333*(up2) + 0.334*(up1) $);
	\coordinate (c4) at ($ 0.333*(mup3) + 0.333*(up2) + 0.334*(up3) $);
	\coordinate (c5) at ($ 0.333*(mup3) + 0.333*(mup4) + 0.334*(up3) $);
	\coordinate (c6) at ($ 0.333*(mup4) + 0.333*(mup5) + 0.334*(up3) $);
	\coordinate (c7) at ($ 0.333*(mup1) + 0.333*(mup2) + 0.334*(mlp2) $);
	\coordinate (c8) at ($ 0.333*(mup2) + 0.333*(mlp2) + 0.334*(mlp3) $);
	\coordinate (c9) at ($ 0.333*(mup2) + 0.333*(mlp3) + 0.334*(mup3) $);
	\coordinate (c10) at ($ 0.333*(mup3) + 0.333*(mlp3) + 0.334*(mup4) $);
	\coordinate (c11) at ($ 0.333*(mlp3) + 0.333*(mlp4) + 0.334*(mup4) $);
	\coordinate (c12) at ($ 0.333*(mup4) + 0.333*(mlp4) + 0.334*(mup5) $);
	\coordinate (c13) at ($ 0.333*(mup1) + 0.333*(lp1) + 0.334*(mlp2) $);
	\coordinate (c14) at ($ 0.333*(mlp2) + 0.333*(lp1) + 0.334*(mlp3) $);
	\coordinate (c15) at ($ 0.333*(lp1) + 0.333*(lp2) + 0.334*(mlp3) $);
	\coordinate (c16) at ($ 0.333*(mlp3) + 0.333*(lp2) + 0.334*(mlp4) $);
	\coordinate (c17) at ($ 0.333*(lp2) + 0.333*(lp3) + 0.334*(mlp4) $);
	\coordinate (c18) at ($ 0.333*(lp3) + 0.333*(mlp4) + 0.334*(mup5) $);
	\draw[black, fill = black!10] (mup1) -- (up1) -- (mup2) -- (mup1);
	\draw[black, fill = black!10] (up1) -- (mup2) -- (mup3) -- (up1);
	\draw[black, fill = black!10] (up1) -- (up2) -- (mup3) -- (up1);
	\draw[black, fill = black!10] (up2) -- (up3) -- (mup3) -- (up2);
	\draw[black, fill = black!10] (up3) -- (mup3) -- (mup4) -- (up3);
	\draw[black, fill = black!10] (up3) -- (mup4) -- (mup5) -- (up3);
	\draw[black, fill = black!10] (mup1) -- (mup2) -- (mlp2) -- (mup1);
	\draw[black, fill = black!10] (mup2) -- (mup3) -- (mlp3) -- (mup2);
	\draw[black, fill = black!10] (mup3) -- (mup4) -- (mlp3) -- (mup3);
	\draw[black, fill = black!10] (mup2) -- (mlp2) -- (mlp3) -- (mup2);
	\draw[black, fill = black!10] (mup4) -- (mlp4) -- (mlp3) -- (mup4);
	\draw[black, fill = black!10] (mup4) -- (mup5) -- (mlp4) -- (mup4);
	\draw[black, fill = black!10] (mup1) -- (lp1) -- (mlp2) -- (mup1);
	\draw[black, fill = black!10] (lp1) -- (mlp2) -- (mlp3) -- (lp1);
	\draw[black, fill = black!10] (lp1) -- (lp2) -- (mlp3) -- (lp1);
	\draw[black, fill = black!10] (lp2) -- (mlp3) -- (mlp4) -- (lp2);
	\draw[black, fill = black!10] (lp2) -- (lp3) -- (mlp4) -- (lp2);
	\draw[black, fill = black!10] (lp3) -- (mlp4) -- (mup5) -- (lp3);
	%
	\begin{scope}[xshift=3.5cm, yshift=-1.5cm, xscale=0.38, yscale=0.66]
		\coordinate (mup1) at (-1.95,0.0);
		\coordinate (mup2) at (-1.1,0.2);
		\coordinate (mup3) at (-0.2,0.4);
		\coordinate (mup4) at (1.1,0.2);
		\coordinate (mup5) at (1.95,0.0);
		\coordinate (up1) at (-1.5,0.65);
		\coordinate (up2) at (0.0,0.95);
		\coordinate (up3) at (1.1,0.8);
		\coordinate (mlp2) at (-1.1,-0.3);
		\coordinate (mlp3) at (-0.2,-0.4);
		\coordinate (mlp4) at (1.0,-0.3);
		\coordinate (lp1) at (-1.5,-0.65);
		\coordinate (lp2) at (0.0,-0.95);
		\coordinate (lp3) at (1.1,-0.8);
		\coordinate (c1) at ($ 0.5*(mup3) + 0.5*(mlp3) $);
		\coordinate (c2) at ($ 0.5*(lp1) + 0.5*(lp2) $);
		\coordinate (c3) at ($ 0.5*(up1) + 0.5*(mup3) $);
		\draw[black, fill = black!10] (mup1) -- (up1) -- (mup2) -- (mup1);
		\draw[black, fill = black!10] (up1) -- (mup2) -- (mup3) -- (up1);
		\draw[black, fill = black!10] (up1) -- (up2) -- (mup3) -- (up1);
		\draw[black, fill = black!10] (up2) -- (up3) -- (mup3) -- (up2);
		\draw[black, fill = black!10] (up3) -- (mup3) -- (mup4) -- (up3);
		\draw[black, fill = black!10] (up3) -- (mup4) -- (mup5) -- (up3);
		\draw[black, fill = black!10] (mup1) -- (mup2) -- (mlp2) -- (mup1);
		\draw[black, fill = black!10] (mup2) -- (mup3) -- (mlp3) -- (mup2);
		\draw[black, fill = black!10] (mup3) -- (mup4) -- (mlp3) -- (mup3);
		\draw[black, fill = black!10] (mup2) -- (mlp2) -- (mlp3) -- (mup2);
		\draw[black, fill = black!10] (mup4) -- (mlp4) -- (mlp3) -- (mup4);
		\draw[black, fill = black!10] (mup4) -- (mup5) -- (mlp4) -- (mup4);
		\draw[black, fill = black!10] (mup1) -- (lp1) -- (mlp2) -- (mup1);
		\draw[black, fill = black!10] (lp1) -- (mlp2) -- (mlp3) -- (lp1);
		\draw[black, fill = black!10] (lp1) -- (lp2) -- (mlp3) -- (lp1);
		\draw[black, fill = black!10] (lp2) -- (mlp3) -- (mlp4) -- (lp2);
		\draw[black, fill = black!10] (lp2) -- (lp3) -- (mlp4) -- (lp2);
		\draw[black, fill = black!10] (lp3) -- (mlp4) -- (mup5) -- (lp3);
		\draw[black, fill = black!10] (up2) -- (mup4);
		\draw[black, fill = black!10] (c1) -- (mup2);
		\draw[black, fill = black!10] (c1) -- (mup4);
		\draw[black, fill = black!10] (c2) -- (mlp3);
		\draw[black, fill = black!10] (mlp3) -- (lp3);
		\draw[black, fill = black!10] (c3) -- (mup2);
		\draw[black, fill = black!10] (c3) -- (up2);
	\end{scope}
	\draw[black, thick] (0.9, 0.5) --+ (1.3,0)node[right] {$\mathcal{V}_1$};
	\draw[black, thick] (-2, -1) --+ (0.4,-0.6)node[below] {$\vect{r}_{l,m,n}^{\,(1)}$} -- (-1,-1);
	\foreach \x in {-2,...,2}
	\foreach \y in {-1,...,1}
	{
		\fill[red!60!black] (\x,\y) circle (1.5pt);
	}
	\draw[black, thick] (3.54, -1.4) --+ (0,0.9)node[above] {$\mathcal{V}_2$};
	\draw[black, thick] (2.78, -2.1) --+ (-0.7,0.2)node[left] {$\vect{r}_{l,m,n}^{\,(2)}$} -- (2.78,-1.28);
	\foreach \x in {0,...,5}
	\foreach \y in {0,...,3}
	{
		\fill[blue!70!black] (2.78+0.3*\x,-2.1+0.41*\y) circle (1.5pt);
	}
\end{tikzpicture}%
%
  \caption{Graphical illustration of the AIM grid used for two objects with different material properties.}%
  \label{fig:objAIMgeom}%
\end{figure}

Each object $i$ is enclosed in an independent regular Cartesian grid, with grid points denoted as $\rvec_{l,m,n}^{\,(i)}$, where $i = 1,2\ldots N_\mathrm{obj}$. An example setup for two objects is shown in \figref{fig:objAIMgeom}.
The number of grid points along each direction, $N_x^{(i)}$, $N_y^{(i)}$ and $N_z^{(i)}$, are chosen such that the grid completely encloses the associated object.
A near-region size is chosen based on mesh dimensions and the object's electrical size at the highest frequency of interest~\cite{AIMbles}.
Details regarding the choice of grid parameters are discussed in \secref{sec:objAIM}.

With the AIM, the interior problem matrices in~\eqref{eq:MFIE1dis} can be written as
\begin{align}
  \Lin &\approx \LinNR + \matr{W}^{(i)}\matr{G}_{i}\matr{P}_{i} + \dfrac{1}{k_i^2}\matr{W}_{i,\phi}\matr{G}_{i,\phi}\matr{P}_{i,\phi} - \Lmat[i,\mathrm{P}],\label{eq:AIM_Lin}\\
  \Kin &\approx \KinNR + \matr{W}_{i}\matr{G}_{i,\nabla}\matr{P}_{i} - \Kmat[i,\mathrm{P}],\label{eq:AIM_Kin}
\end{align}
where $\LinNR$ and $\KinNR$ are sparse and contain the near-region entries of $\Lin$ and $\Kin$, respectively. These matrices are computed accurately with direct integration.
Matrices $\matr{P}$ and $\matr{P}_\phi$ are, respectively, vector and scalar projection matrices which project sources from the mesh to the AIM grid, while $\matr{G}$, $\matr{G}_\phi$ and $\matr{G}_{\nabla}$ are convolution matrices which transform grid sources to grid potentials.
Matrices $\matr{G}$ and $\matr{G}_\phi$ encode $G\rrpki$, while $\matr{G}_{\nabla}$ encodes $\nabla G\rrpki$.
The grid potentials are then interpolated back on the original mesh with vector and scalar interpolation matrices $\matr{W}$ and $\matr{W}_\phi$.
Due to the singularity of $G\rrpki$ when $\vect{r} = \vecprime{r}$, the direct integration-based entries of $\LinNR$ and $\KinNR$ are used in the near-region, instead of the interpolation-based grid potentials.
Therefore, a key step in the AIM is to use the precorrection matrices $\Lmat[i,\mathrm{P}]$ and $\Kmat[i,\mathrm{P}]$ to cancel out the grid potentials in the near-region.
Matrix-vector products involving $\matr{G}$, $\matr{G}_\phi$ and $\matr{G}_{\nabla}$ are computed in the spectral domain with a 3D Fast Fourier Transform (FFT)~\cite{pfftmain}.
Details on the entries of the projection, convolution and interpolation matrices can be found in literature~\cite{pfftmain}.
In the following, we define for brevity
\begin{align}
  \LinNRP &\triangleq \LinNR - \Lmat[i,\mathrm{P}],\label{eq:LinNRP}\\
  \KinNRP &\triangleq \KinNR - \Kmat[i,\mathrm{P}],\label{eq:KinNRP}.
\end{align}
Substituting~\eqref{eq:AIM_Lin}--\eqref{eq:KinNRP} in~\eqref{eq:GIBCdef}, we obtain
\begin{multline}
  \matr{Z}_{i} = \dfrac{-1}{j\omega\epsilon_i} {\left( \LinNRP + \matr{W}_{i}\matr{G}_{i}\matr{P}_{i} +  \dfrac{1}{k_i^2}\matr{W}_{i,\phi}\matr{G}_{i,\phi}\matr{P}_{i,\phi} \right)}^{-1}\cdot\\ \left( \KPxinNRP + \matr{W}_{i}\matr{G}_{i,\nabla}\matr{P}_{i} \right), \label{eq:GIBCAIM}
\end{multline}
for $i = 1,2\ldots N_\mathrm{obj}$.

At iteration $k$ of the iterative solution of~\eqref{eq:GIBCsys}, which we refer to as the ``outer'' iteration, the matrix-vector product $\matr{Z}\matr{H}^{(k)}$ is obtained by first computing
\begin{align}
  \matr{a}^{(k)} \triangleq \left( \KPxinNRP + \matr{W}_{i}\matr{G}_{i,\nabla}\matr{P}_{i} \right) {\Hmat[i]}^{(k)},\label{eq:nest02}
\end{align}
and then solving the system
\begin{multline}
  -j\omega\epsilon_i \left( \LinNRP + \matr{W}_{i}\matr{G}_{i}\matr{P}_{i} + \dfrac{1}{k_i^2}\matr{W}_{i,\phi}\matr{G}_{i\phi}\matr{P}_{i\phi} \right) \cdot\\ \left( \matr{Z}_i\matr{H}_i^{(k)} \right) = \matr{a}^{(k)},\label{eq:nest2}
\end{multline}
for $\matr{Z}_i\matr{H}_i^{(k)}$ with an iterative method at each $k$.
Equations~\eqref{eq:nest02} and~\eqref{eq:nest2} define a generalized impedance boundary condition where the interior problem matrices are expressed with an AIM-based approximation, and their products with vectors can be calculated on the fly in a scalable way, for both small and large objects.
The accuracy of the approximation is controlled via the AIM grid parameters such as spacing and interpolation order, as well as the convergence tolerance for the iterative solution of~\eqref{eq:nest2}.
Therefore, it is unnecessary to build, store and factorize any dense matrices, leading to significant computational savings, both in terms of CPU time and memory.
The efficient solution of~\eqref{eq:nest2} is discussed next.

\subsubsection{Preconditioning the Nested System}\label{sec:precond}

The number of iterations required to solve the nested system~\eqref{eq:nest2} may significantly influence the overall matrix-vector product cost for each outer iteration $k$.
Minimizing this cost requires an effective preconditioner.
We propose to use the following sparse near-region preconditioner $\matr{M}$ for the system in~\eqref{eq:nest2}~\cite{near_zone_PC}:
\begin{align}
  \matr{M}_i = -j\omega\epsilon_i \LinNR. \label{eq:nest2pc}
\end{align}
Preconditioner $\matr{M}_i$ can be factorized efficiently in advance for each object, using a sparse LU factorization.
The proposed preconditioner has the following advantages:

\begin{enumerate}[label=\alph*)]
  
\item For conductors, as skin effect develops with increasing frequency, $\Lin$ becomes increasingly sparse and diagonally dominant.
  Thus $\LinNR \approx \Lin$, which leads to extremely quick convergence of the nested solution of~\eqref{eq:nest2}.
  In particular, when skin depth becomes smaller than the near-region size, then $\LinNR = \Lin$ and the factorization of $\matr{M}_i$ effectively leads to a direct solution of~\eqref{eq:nest2}, requiring one iteration to converge.
  Thus, the proposed preconditioner is naturally able to take advantage of the fast decay of the conductive medium Green's function.
  
\item For small objects, it is more efficient to directly factorize $\Lin$ rather than solve~\eqref{eq:nest2} iteratively.
  This is naturally achieved by the proposed preconditioner, without the need to set a heuristic threshold to determine which objects are considered small or large.
  For objects whose bounding box is smaller than the near-region size, $\matr{M}_i = -j\omega\epsilon_i \Lin$.
  Therefore, factorizing $\matr{M}_i$ effectively leads to a direct solution of~\eqref{eq:nest2}, and AIM is automatically invoked only for objects large enough to require it.

\item In the general case of dielectric objects or conductors at low frequency, it has been shown that the near-region preconditioner~\eqref{eq:nest2pc} regularizes the EFIE operator~\cite{near_zone_PC}, which is precisely the operator in~\eqref{eq:nest2}.
The regularization property ensures quick convergence of the nested system~\eqref{eq:nest2}, as shown for realistic examples in \secref{sec:results}.
  
\end{enumerate}
The features listed above ensure that the nested solution of~\eqref{eq:nest2} is performed efficiently at each outer iteration $k$, for both dielectrics and conductors, without the need to expand the list of unknowns, as in the eAEFIE~\cite{eaefie01}.

\subsection{Final System of Equations}\label{sec:system}
The final system of equations can now be expressed using the proposed expressions for the surface impedance operator.
To overcome the low frequency breakdown of $\Lout$ in~\eqref{eq:EFIE01dis}, the augmented EFIE~\cite{aefie2,aefie1} is used whereby the surface charge density $\rho_s$ is also taken as unknown.
Equations~\eqref{eq:GIBCAIM},~\eqref{eq:glGIBCdef} and~\eqref{eq:GIBCdef} can be used in the augmented form of~\eqref{eq:EFIE01dis} to obtain the final system,
\begin{align}
  \begin{bmatrix}
    jk_0\, \Lout[A] - \left( \KPxout \right)\matr{Z} & \matr{D}^T \Lout[\phi] \matr{B} \\
    \matr{F} \matr{D} & jk_0 \matr{I}
  \end{bmatrix}
  \begin{bmatrix} \Hmat \\ c_0\rhomat \end{bmatrix}
  =
  \begin{bmatrix} -\frac{\Emat[\mathrm{inc}]}{\eta_0} \\ \matr{0} \end{bmatrix}, \label{eq:sys}
\end{align}
where $\Lout[A]$ and $\Lout[\phi]$ are the vector and scalar potential parts of $\Lout$, respectively.
Matrix $\matr{I}$ is the identity, $c_0$ is the speed of light in $\mathcal{V}_0$, and definitions of sparse matrices $\matr{D}$, $\matr{B}$ and $\matr{F}$ can be found in literature~\cite{aefie2}.
Column vector $\rhomat$ stores the unknown coefficients of $\rho_s$, which is expanded with pulse basis functions.

Matrix-vector products involving $\Lout[A]$, $\Lout[\phi]$ and $\Kout$ are also accelerated with the AIM by enclosing the entire structure in a global regular grid, based on the electrical properties of $\mathcal{V}_0$.
The constraint preconditioner proposed previously~\cite{aefie2} is used to improve convergence of the iterative solution of~\eqref{eq:sys}.
Note that since dual basis functions are not used in the proposed formulation, the term $\KPxout$ in \eqref{eq:sys} is not diagonally dominant.
Thus, we use only $jk_0\,\mathrm{diag}(\Lout[A])$ in the (1,1) block of the constraint preconditioner~\cite{aefie2}.
The proposed method ensures that every matrix-vector product in the iterative solution of \eqref{eq:sys} is accelerated without the need to take $\Emat$ as unknown, unlike in the eAEFIE~\cite{eaefie01}.

An important feature of the proposed method is that it is amenable to parallelization on large compute clusters.
The distributed-memory parallelization of the AIM has recently garnered significant interest~\cite{AIMMPI0,AIMMPI1,EUCAP2020} and has been achieved with good efficiency.
Several open-source libraries exist which allow flexible distributed-memory parallelization of FFTs~\cite{FFTW,PFFT}.
In addition, the memory available on individual compute nodes in a cluster is a fixed constraint, hence the memory saved by the use of AIM in the interior problem is extremely important when objects are large.
Therefore, the proposed method seems to be well suited for distributed-memory parallelization.

The matrix-vector products involved in iteratively solving~\eqref{eq:nest2} may also be accelerated by using the multi-level fast multipole algorithm~(MLFMA)~\cite{MLFMA}, instead of the AIM.
Unlike the AIM, the MLFMA avoids a volumetric grid, and may be well suited when an object occupies a large volume and its mesh is sparse, such as a large sphere.
In this work, the AIM was chosen for the following reasons:
\begin{itemize}
\item The AIM is applicable at low frequencies, or equivalently for sub-wavelength objects, without special treatment.
The MLFMA instead must take on a different form to handle the low frequency case.
This is particularly important for on-chip applications, where individual objects, such as microvias, may be extremely sub-wavelength in size.
The mixed-form fast multipole algorithm~\cite{MFFMA} was proposed to model both low and high frequency physics, and would also be an interesting candidate to accelerate matrix-vector products in the interior problem.
\item The use of AIM in both the exterior and interior problems makes the code easier to parallelize, particularly on distributed memory systems.
Since the matrix-vector products in the interior problem are nested within those of the exterior problem, the same mesh and grid point distribution strategy may be reused, potentially minimizing communication between nodes as well as memory usage.
This may be more challenging to achieve in a mixed code that interweaves both the AIM and the MLFMA.
\end{itemize}

\section{Accuracy of the AIM for the Interior Problem} \label{sec:objAIM}

\begin{figure}[t]
  \centering
  \includegraphics[width=\linewidth,trim={0 20 0 20},clip=true]{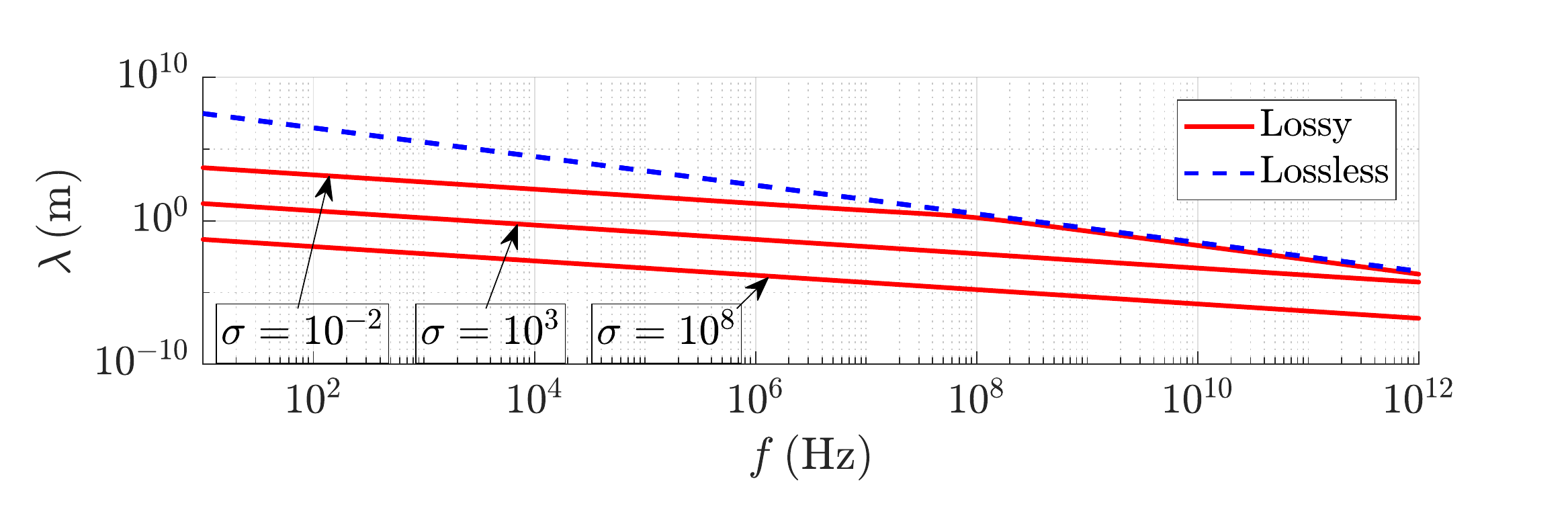}\\
  \includegraphics[width=\linewidth,trim={0 20 0 20},clip=true]{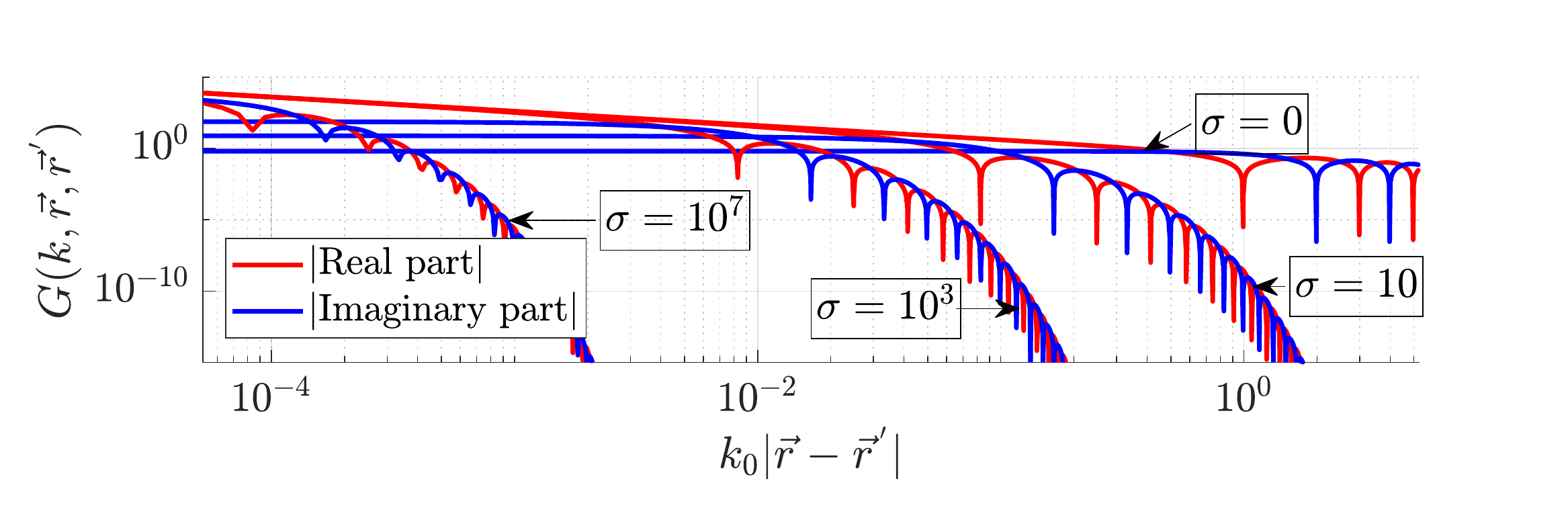}
  \caption{Top panel: wavelength in lossy media. Bottom panel: homogeneous
    Green's function for lossy media. Conductivity is reported S/m. Unit of $\Grrpk$ is $1/$m.}
  \label{fig:lossy_figs}
\end{figure}

An important consideration in the proposed method is the accuracy of AIM when applied to media with arbitrary electrical properties, particularly conductors.
To the best of our knowledge, the use of AIM has so far been restricted to lossless or mildly lossy materials~\cite{AIM_multiple_grid,pfftcomposite}.
In order to extend the use of AIM to highly lossy media, where the wavelength is extremely small compared to the lossless case, a detailed error analysis is necessary, and has not been reported previously.
In this section, a generalized error analysis of AIM approximations is presented, parameterized in terms of losses in the associated medium.
This analysis will provide us with the appropriate choice of AIM grid parameters to ensure both accuracy and efficiency.

Existing error analyses on the AIM assume a grid spacing based on the wavelength in the relevant medium~\cite{AIMbles,pffterrors,pfftcompare}, typically chosen as $\lambda_i/10$ or $\lambda_i/20$, where $\lambda_i$ is the wavelength inside the $i^\mathrm{th}$ object~\cite{AIMbles}.
However, this choice may not be appropriate for conductive media because:
\begin{itemize}

\item  As conductivity increases, the wavelength in the medium shrinks significantly.
  This implies that an extremely fine AIM grid may be required to capture the rapid field variations, leading to very expensive FFTs.
  The real part of wavelengths for materials with different values of conductivity are shown in the top panel of \figref{fig:lossy_figs}, which indicates that lossy media may require a grid spacing several orders of magnitude finer than lossless media.
  
\item At the same time, the Green's function for conductive media decays significantly faster than in the lossless case, and therefore has less of an impact on accuracy in the far-region.
  This decay is not taken into account when grid parameters are chosen based only on the wavelength in the medium, as in conventional AIM implementations~\cite{AIMbles} and error analyses~\cite{pffterrors,pfftcompare}.
  This is visualized in the bottom panel of \figref{fig:lossy_figs}, where the Green's function for various values of $\sigma_i$ is plotted.
  At sufficiently high frequency, the exponential decay of the Green's function ensures that far-away interactions are extremely weak and need not be modeled accurately.
  However, at low and intermediate frequencies, when the oscillations are faster but the Green's function has not decayed significantly, the AIM must still be able to model the Green's function accurately.

\end{itemize}

The efficiency of AIM for lossy media is therefore not obvious, and the optimal choice of grid spacing must be investigated in the context of losses.
In this section, we provide an error analysis of the AIM which explicitly takes into account the decay of the Green's function due to losses, and thereby provides insight into the applicability of AIM for conductive media.
%

It has been shown for scalar basis functions that when moment-matching techniques~\cite{pfftcompare} are used in far-region computations, the accuracy of AIM depends on its ability to approximate the Green’s function with an interpolating polynomial~\cite{pfftmain}.
The same can be shown for vectorial basis functions.
Since the homogeneous Green's function is reciprocal,
\begin{align}
  \Grrpki = \green{G(k_i, \vect{r}^{\,'}, \vect{r})},\quad i = 1,2 \ldots N_\mathrm{obj}, \label{eq:Grecip}
\end{align}
we only need to study the interpolation error of $\Grrpki$ either
on the source stencil or on the test stencil.
Without loss of generality, we can set $\vect{r}^{\,'} = 0$, so that
${r \triangleq |\vect{r} - \vect{r}^{\,'}| = \abs{\vect{r}}}$ and ${\Grrpki = \Grki}$.
To parameterize losses, we can define $s$, a unit-less, normalized loss factor, as
${s \triangleq \frac{\sigma}{\omega\epsilon_i}}$,
so that 
\begin{align}
  k = k_{r,i} \sqrt{1 - js}, \label{eq:kdefs}
\end{align}
where $k_{r,i} = k_0\sqrt{\epsilon_{r,i}\mu_{r,i}}$, and $\epsilon_{r,i}$ and $\mu_{r,i}$ are the relative permittivity and permeability of the $i^{\mathrm{th}}$ object.
Furthermore, we normalize any physical length, say $d$, as
${\theta_{d,i} \triangleq k_{r,i}d}$.

Consider a regular AIM grid with a stencil which has $n + 1$ points in each direction.
Given $n+1$ samples $(r_0, \green{G(k_i,r_0)}), (r_1, \green{G(k_i,r_1)}) \ldots (r_n, \green{G(k_i,r_n)})$, the Lagrange polynomial approximation of $\Grki$ can be written as~\cite{abrstegun}
\begin{align}
  \Grki[a] = \sum_{q=0}^n \green{G(k_i,r_q)} \left( \prod_{0 \leq m \leq n}^{m \neq q} \dfrac{r - r_m}{r_q - r_m} \right),
\end{align}
and the goal is to study the remainder
\begin{align}
  \Delta G_n = \abs{\Grki - \Grki[a]},
\end{align}
which represents the interpolation error for order $n$.
Written in terms of normalized physical lengths, $\Delta G_n$ can be bound as~\cite{abrstegun}
\begin{align}
  \dfrac{\Delta G_n}{k_{r,i}} \leq \dfrac{{\left(\theta_n - \theta_0\right)}^{n+1}}{\left(n+1\right)!}\, \max_{\theta_0 \leq \theta_\xi \leq \theta_n} \abs{ \dfrac{d^{\,n+1}}{d\theta_\xi^{\,n+1}}\left[ \dfrac{e^{-j\theta_\xi \sqrt{1 - js}}}{\theta_\xi} \right] },\label{eq:Lbounds}
\end{align}
where $\theta_0$ and $\theta_n$ are the stencil coordinates nearest to and farthest from the source point, respectively.
Using the product rule for the $n^{\mathrm{th}}$ order derivative,
\begin{align}
  \dfrac{d^{\,n}}{dt^{\,n}}\left[u(t)v(t)\right] = \sum_{l = 0}^{n}
  \begin{pmatrix}
    n \\ l
  \end{pmatrix}
  \dfrac{d^{\,n-l}u(t)}{dt^{\,n-l}} \dfrac{d^{\,l}v(t)}{dt^{\,l}},
\end{align}
equation~\eqref{eq:Lbounds} can be simplified as
\begin{align}
  \dfrac{\Delta G_n}{k_{r,i}} \leq \dfrac{{\left(\theta_n - \theta_0\right)}^{n+1}}{\left(n+1\right)!}\, \max_{\theta_0 \leq \theta_\xi \leq \theta_n} \abs{ D_n(\theta_\xi, s) }, \label{eq:LboundsD}
\end{align}
where
\begin{multline}
  D_n(\theta_\xi, s) = e^{-j\theta_\xi \sqrt{1 - js}}\,\sum_{l = 0}^{n+1}
  \begin{pmatrix}
    n+1 \\ l
  \end{pmatrix}{\left(-1\right)}^{n+1-l}\cdot\\
  \left(n+1-l\right)! {\left(\dfrac{1}{\theta_\xi}\right)}^{n-l+2}
  {\left(-j\right)}^l {\left(\sqrt{1 - js}\right)}^l. \label{eq:Ddef}
\end{multline}
Since $\abs{D_n(\theta_\xi, s)}$ is a monotonically decreasing function of $\theta_\xi$, $\max_{\theta_0 \leq \theta_\xi \leq \theta_n} \abs{ D_n(\theta_\xi, s) }$ occurs at $\theta_\xi$ = $\theta_0$.
Therefore,~\eqref{eq:LboundsD} can be further simplified as
\begin{align}
  \dfrac{\Delta G_n}{k_{r,i}} \leq \dfrac{{\left(\theta_n - \theta_0\right)}^{n+1}}{\left(n+1\right)!}\, \abs{ D_n(\theta_0, s) }. \label{eq:LboundsD2}
\end{align}
Equation~\eqref{eq:LboundsD2} provides a general error bound for the moment-matching step of the AIM for arbitrary media, parameterized in terms of losses.

\begin{figure}[t]
  \includegraphics[width=\linewidth,trim={0 0 0 0},clip=true]{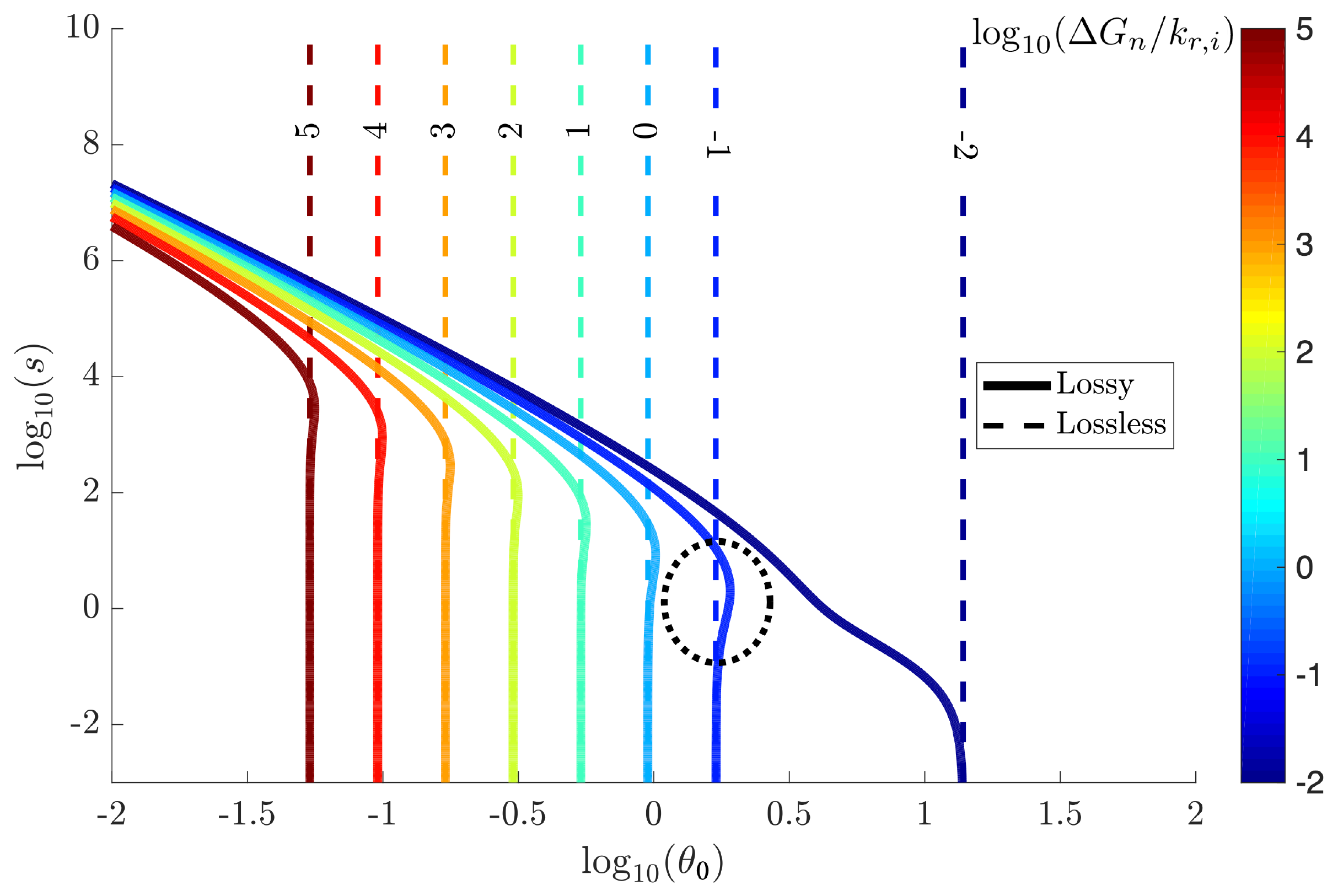}
  \caption{Contour plot of interpolation error bounds, on a logarithmic scale, for the Green's function of lossy and lossless media, as a function of distance ($\theta_0$) and losses ($s$), for ${n=2}$.}\label{fig:R}
\end{figure}

\figref{fig:R} shows a contour plot of $\Delta G_n/k_{r,i}$ as a function of $\theta_0$ and $s$, for ${n=2}$.
The dashed lines represent the lossless case, $\Delta G_{n,\mathrm{lossless}}/k_{r,i}$ where we have set ${s = 0}$, which allows us to compare the interpolation error for lossy and lossless media for the same AIM grid.

The ratio $\Delta G_n/\Delta G_{n,\mathrm{lossless}}$ quantifies the increase in the error bound caused by losses.
Taking into account the wide range of $\theta_0$ and $s$ considered in \figref{fig:R}, we can find that
\begin{align}
  \max_{\theta_0,s} \dfrac{\Delta G_n}{\Delta G_{n,\mathrm{lossless}}} = 1.513,
\end{align}
which occurs for ${\theta_0 = 1.653}$ and ${s = 2.894}$.
Therefore, in the worst case, the error bound for lossy media is only $1.5\times$ larger than the corresponding bound for the lossless case, for the same AIM grid.
For the vast majority of cases, the error bound for the lossy case is comparable to or smaller than the lossless case.
The dotted ellipse in \figref{fig:R} indicates an example region in which the error bound for lossy media is larger than the corresponding bound for the lossless case.
This corresponds to the situations discussed above, in which the oscillations of the Green's function are faster, and its magnitude is large enough that far-region interactions must still be taken into account.

The above analysis leads to the following key conclusion: the same AIM grid can be used for lossy media as for lossless media, with a minimal compromise in accuracy.
The grid size can be chosen as $\lambda_{r,i}/20$ rather than $\lambda_i/20$, where $\lambda_{r,i} = 2\pi/k_{r,i}$.
Even if stricter control over the accuracy is required, the grid only needs to be further refined by a small factor of at most $1.5\times$, rather than several orders of magnitude, as predicted by a wavelength-based grid size~\cite{pfftthesis}.

Next, the above analysis is verified with numerical examples.
Matrices $\Lin$ and $\Kin$ are explicitly generated using AIM for representative structures, and compared with those generated directly without AIM.
The relative error between a matrix $\matr{A}$ (which may represent $\Lin$ or $\Kin$) and its AIM-based counterpart $\matr{A}_\mathrm{AIM}$ is defined as
\begin{align}
  \kappa = \dfrac{\max \abs{\matr{A} - \matr{A}_\mathrm{AIM}}}{\max \abs{\matr{A}}}.
\end{align}

\begin{figure}[t]
  \subfloat[][]
  {
    \includegraphics[width=.5\linewidth,trim={0 0 0 0},clip=true]{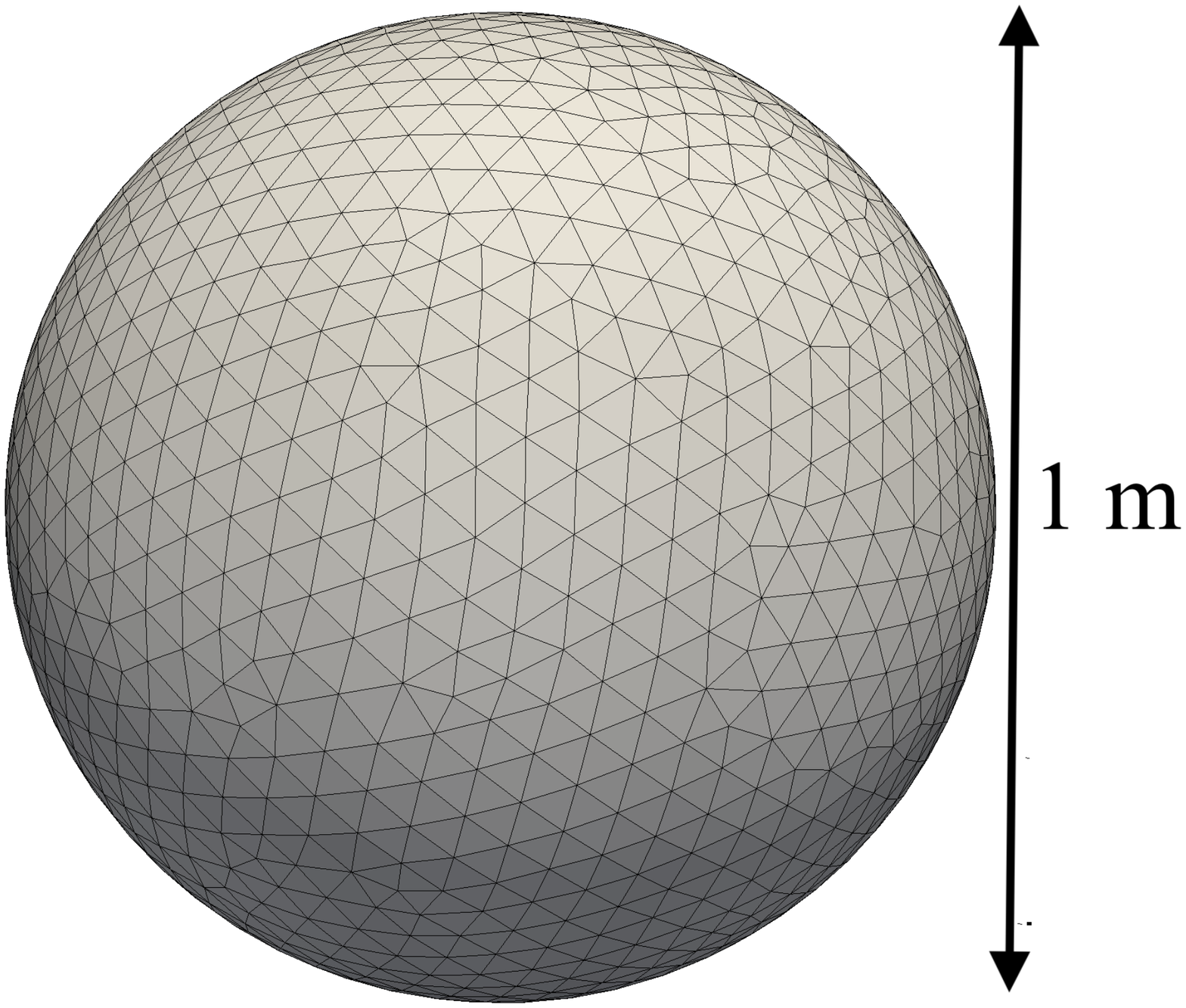}\label{fig:geom:sphere_1}
  }
  \subfloat[][]
  {
    \includegraphics[width=.5\linewidth]{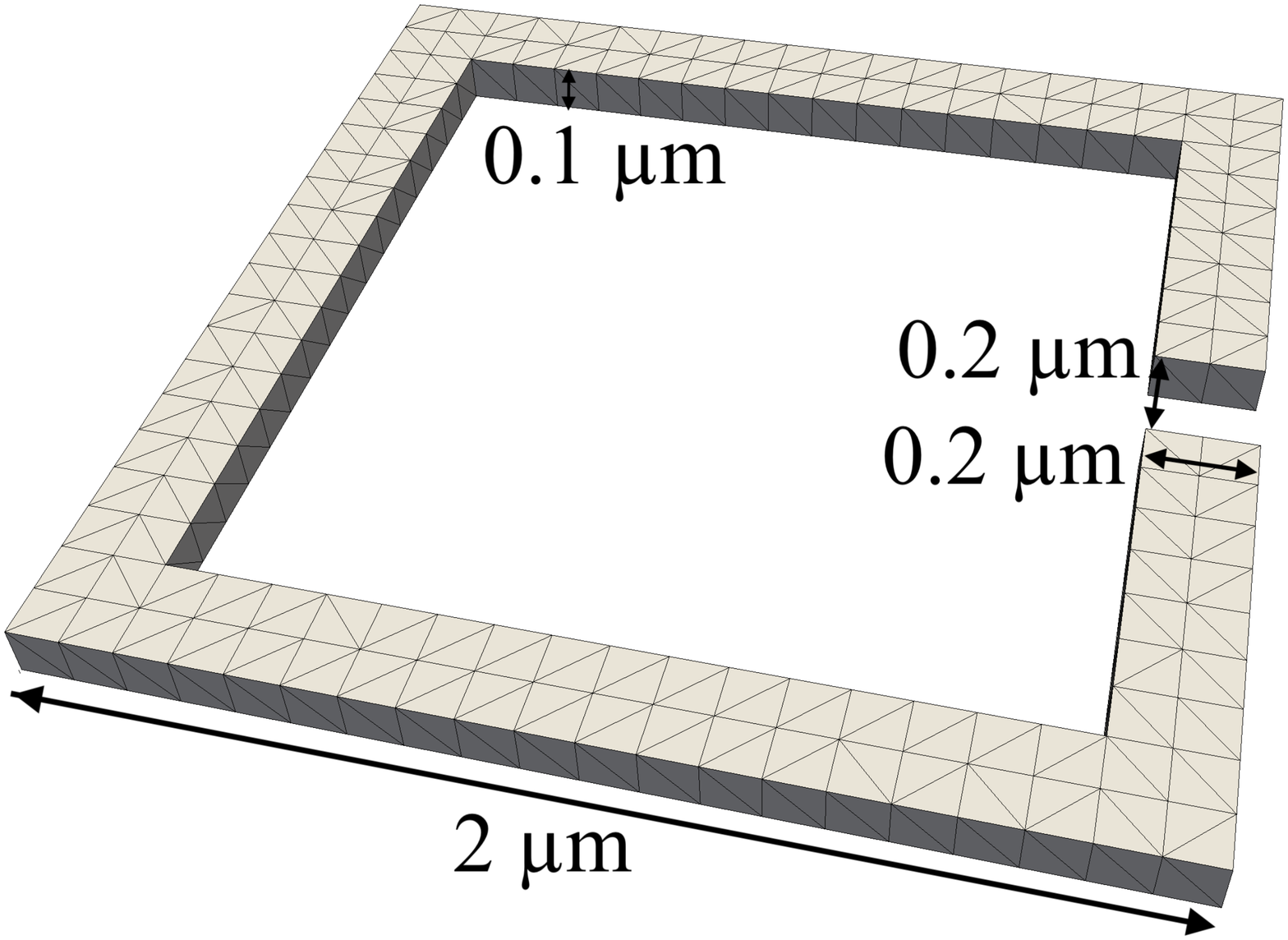}\label{fig:geom:SRR_1}
  }
  \caption{Geometry and mesh of test cases to validate AIM accuracy for lossy media, (a) for the sphere in \secref{sec:results:accAIM:sphere}, and (b) for the split ring resonator in \secref{sec:results:accAIM:SRR}.}
\end{figure}

\subsubsection{Sphere}\label{sec:results:accAIM:sphere}

\begin{figure}[t]
  \includegraphics[width=\linewidth]{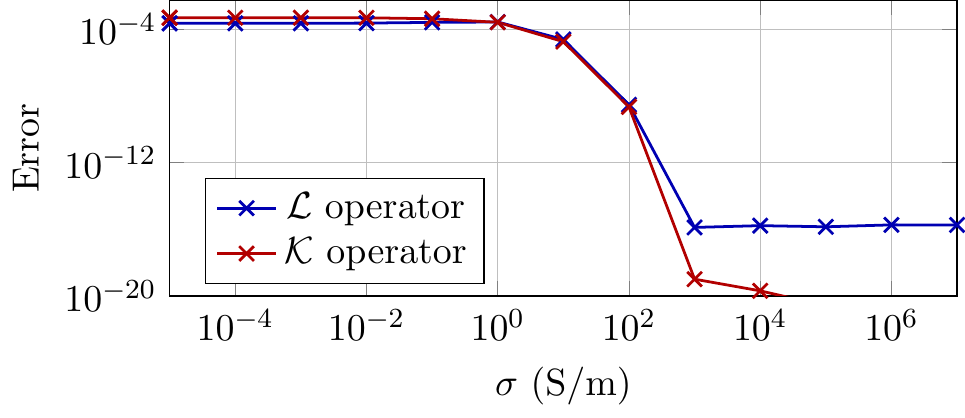}\\
  \includegraphics[width=\linewidth]{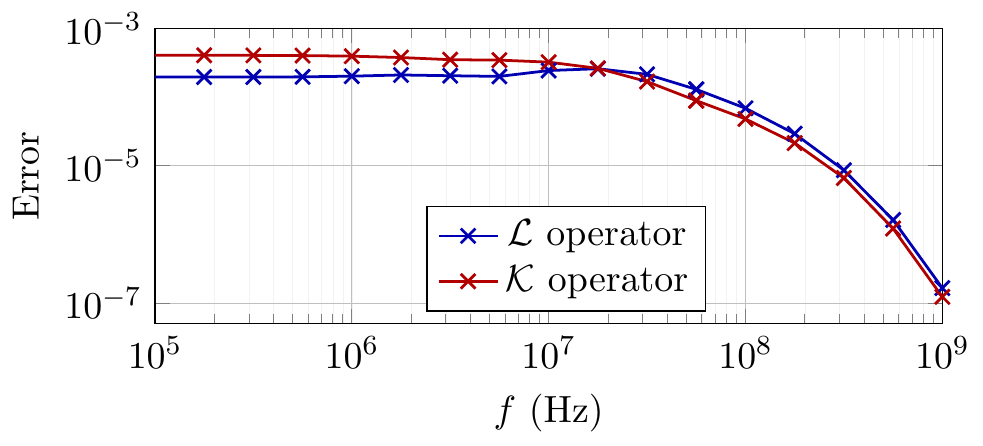}
  \caption{Relative error in AIM-based matrices for the interior problem, for the sphere in \secref{sec:results:accAIM:sphere}, as a function of conductivity (top panel), and frequency (bottom panel).}\label{fig:AIMErrSphere}
\end{figure}

First, we consider a sphere of diameter $1\,$m with ${\epsilon_r = 2.5}$, discretized with 3,168 triangular elements as shown in \figref{fig:geom:sphere_1}.
An AIM grid of ${30 \times 30 \times 30}$ points was used, with a near-region diameter of 11 grid points.
The top panel of \figref{fig:AIMErrSphere} shows the relative error $\kappa$ for the discretized $\opL$ and $\opK$ operators at $200\,$MHz over a very wide range of conductivity, $\sigma \in \left[10^{-7},\, 10^{7}\right]\,$S/m.
The error is below $10^{-3}$ and near $10^{-4}$ across the entire range of conductivities considered, from dielectric to near-PEC regimes.
For extremely conductive media, the sharp decrease in error confirms that the primary contribution to the matrices is due to near-region interactions, which are computed accurately with the MoM.

The bottom panel of \figref{fig:AIMErrSphere} shows the relative error $\kappa$ for a fixed conductivity of $10\,$S/m, and frequencies in the range $\left[0.1\,\text{MHz}, 1\,\text{GHz}\right]$ which corresponds to skin depths in the range $\left[0.5\,\text{m}, 5\,\text{mm}\right]$.
Again, the error is below $10^{-3}$ for the entire range of frequency.

\subsubsection{Split Ring Resonator}\label{sec:results:accAIM:SRR}

\begin{figure}[t]
  \includegraphics[width=\linewidth]{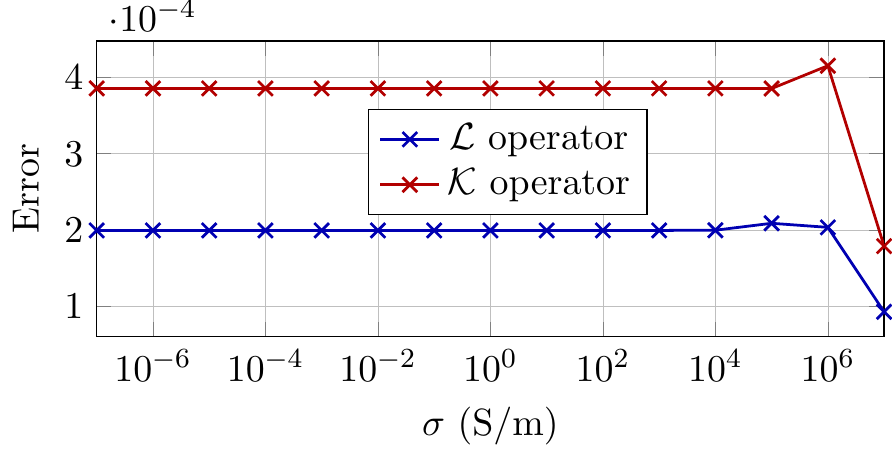}\\
  \includegraphics[width=\linewidth]{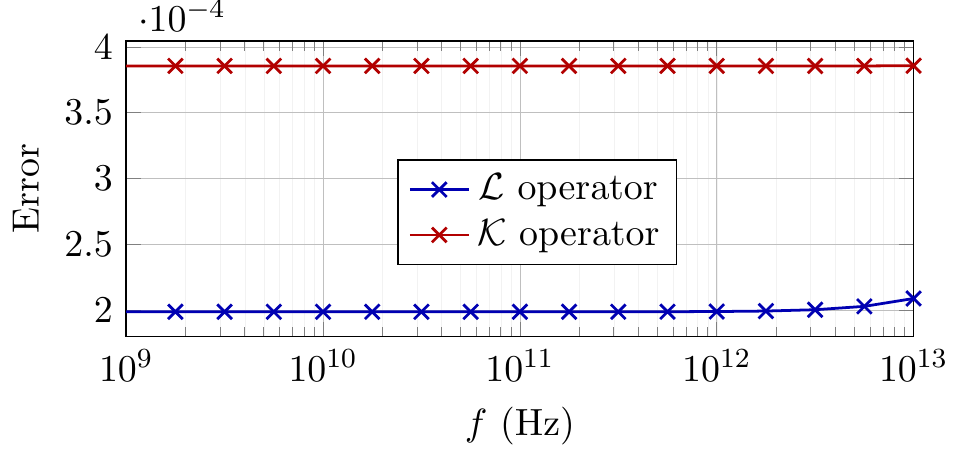}
  \caption{Relative error in AIM-based matrices for the interior problem, for the SRR in \secref{sec:results:accAIM:SRR}, as a function of conductivity (top panel), and frequency (bottom panel).}\label{fig:AIMErrSRR}
\end{figure}

Next, we consider a square split ring resonator (SRR), shown in \figref{fig:geom:SRR_1}, with side length $2\,\mu\mathrm{m}$, gap width $0.2\,\mu\mathrm{m}$, thickness $0.1\,\mu\mathrm{m}$, and $\epsilon_r = 11$.
The SRR is meshed with 848 triangles.
An AIM grid of $20 \times 20 \times 3$ points was used, with a near-region diameter of 11 grid points.
SRR-based structures are of practical importance as unit cells in metamaterial applications, where accurately accounting for skin depth is essential to account for losses~\cite{SRRmetaCond}.
Structures with a similar geometry are also relevant in dielectric metasurfaces~\cite{SRRmetaDiel}, as well as on-chip components such as inductor coils~\cite{fastmaxwell}.
The top panel of \figref{fig:AIMErrSRR} shows relative errors $\kappa$ in AIM-based matrices $\Lin$ and $\Kin$ for conductivies ranging from dielectric to near-PEC regimes, at $10\,$THz.
The bottom panel of \figref{fig:AIMErrSRR} shows $\kappa$ for $\Lin$ and $\Kin$ for a fixed conductivity of $10^{5}\,$S/m, and frequencies such that the skin depth lies in the range $\left[0.5,\, 50\right]\,\mu$m.
The relative error is near or below $10^{-4}$ in all cases.

The numerical results presented above confirm that AIM can be used in the interior problem in a unified manner for a variety of structures and material parameters, ranging from dielectrics to good conductors, as well as over a wide frequency range.


\section{Results}\label{sec:results}

Numerical results are presented in this section which demonstrate the performance of the proposed method for a variety of test cases.
All simulations were performed single-threaded on a 3\,GHz Intel Xeon CPU.
We used PETSc~\cite{petsc-web-page} for sparse matrix manipulation.
The GMRES iterative solver~\cite{gmres} available through PETSc was used for solving~\eqref{eq:nest2} and~\eqref{eq:sys}.
A relative residual norm of $10^{-4}$ was used as the GMRES convergence tolerance for~\eqref{eq:sys}, and a tolerance of $10^{-6}$ was used for solving~\eqref{eq:nest2}.

In all test cases, the proposed method is compared to three existing techniques:
\begin{itemize}
	\item The original GIBC formulation~\cite{GIBC}, where the dense interior problem matrices in~\eqref{eq:GIBCdef} and~\eqref{eq:GIBCsys} are directly assembled and factorized.
	\item A simplified eAEFIE formulation~\eqref{eq:eAEFIEsys} and associated preconditioner~\cite{eaefie01,eaefie02}, which differs from the original eAEFIE as follows:
	\begin{itemize}
		\item Dual basis functions are not employed, since the goal is to demonstrate that the proposed method does not rely on them for convergence over wide ranges of frequency and conductivity, whereas the original eAEFIE does.
		\item As discussed in \secref{sec:existingeAEFIE}, the interior problem can be formulated in terms of the EFIE, the MFIE, or a combination of the two.
		We found that the MFIE leads to the best overall performance and convergence, when dual basis functions are not used.
		\item When solving~\eqref{eq:eAEFIEsys} iteratively, the matrix-vector products are accelerated with the AIM rather than the fast multipole method~\cite{MLFMA}, to facilitate the incorporation of the multilayer Green's function, given that many practical designs involve conductive objects embedded in stratified media.
	\end{itemize}
	Results corresponding to the simplified eAEFIE are denoted by the legend entry ``Simpl.\,eAEFIE''.
	We emphasize that the results presented for this simplified eAEFIE are not representative of the efficiency of the original eAEFIE, but are used to demonstrate that the original eAEFIE heavily relies on dual basis functions in order to achieve convergence within a reasonable number of iterations.
	The numbers of iterations in the SIBC and the simplified eAEFIE, respectively, may be taken as the lower and upper bounds for the number of iterations in the original eAEFIE.
	Even if the lower bound is assumed, the results below indicate that the proposed preconditioning scheme ensures that the time per iteration in the proposed method is sufficiently low, so that the overall solution time would be comparable with that of the original eAEFIE.
	\item The SIBC~\cite{SIBC}, which is inaccurate for dielectrics and for conductors at low and intermediate frequencies, when the skin effect has not fully developed.
	The SIBC is a useful benchmark for performance and convergence, and is accurate for conductive media at high frequencies, when skin effect is fully developed.
\end{itemize}

In keeping with the practice in literature~\cite{GIBC,eaefie02}, a skin depth threshold is employed in the interior problem for the proposed, GIBC, and simplified eAEFIE formulations.
The threshold is chosen such that source-test interactions separated by a distance greater than five skin depths are ignored.
This leads to sparse matrices when the conductivity and frequency are sufficiently large.

\subsection{Sphere: Accuracy Validation}\label{sec:results:sphere_1}

\begin{figure}
	\centering
	\includegraphics[width=\linewidth]{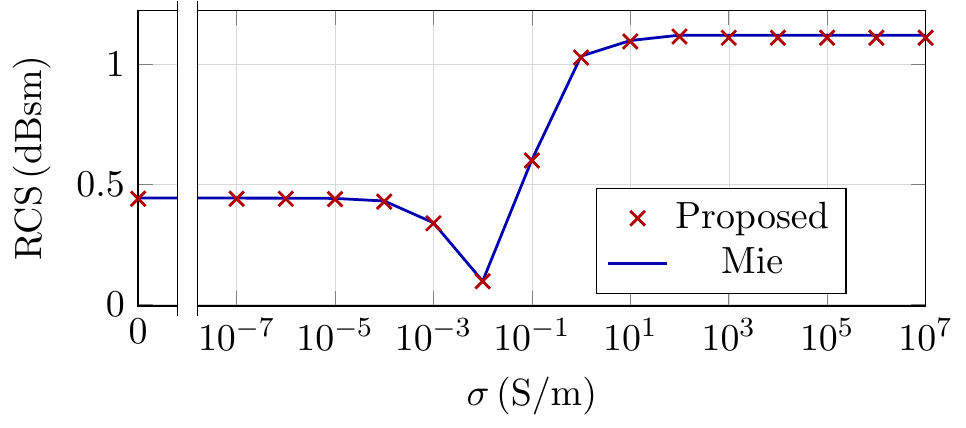}\\
	\includegraphics[width=\linewidth]{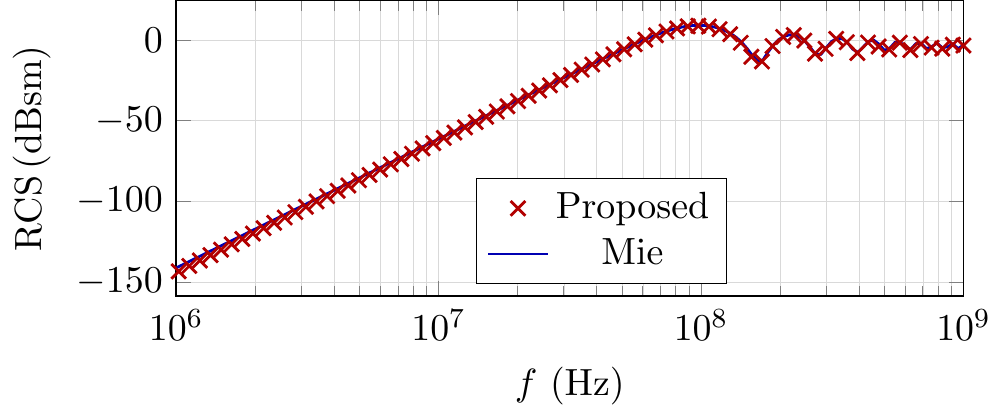}
	\caption{Accuracy validation of the proposed method for the sphere in \secref{sec:results:sphere_1} by comparing the RCS over a range of conductivity (top panel), and frequency (bottom panel).}\label{fig:sphere1mRCS}
\end{figure}

To validate the accuracy of the proposed method, the sphere of \secref{sec:results:accAIM:sphere} is excited with a plane wave at $200\,$MHz.
The monostatic radar cross section (RCS) is computed and compared with the analytical solution obtained via Mie series~\cite{gibson}.
The top panel of \figref{fig:sphere1mRCS} indicates an excellent agreement over a very wide range of conductivities that stretches from perfect dielectric to near-PEC regimes.
The bottom panel of \figref{fig:sphere1mRCS} confirms the excellent agreement also over a wide frequency band covering the entire skin depth transition from near-uniform current distribution to near-PEC behaviour.

\begin{figure}[t]
	\centering
	\subfloat[][]
	{
		\includegraphics[width=\linewidth,trim={0 12mm 0 14mm},clip=false]{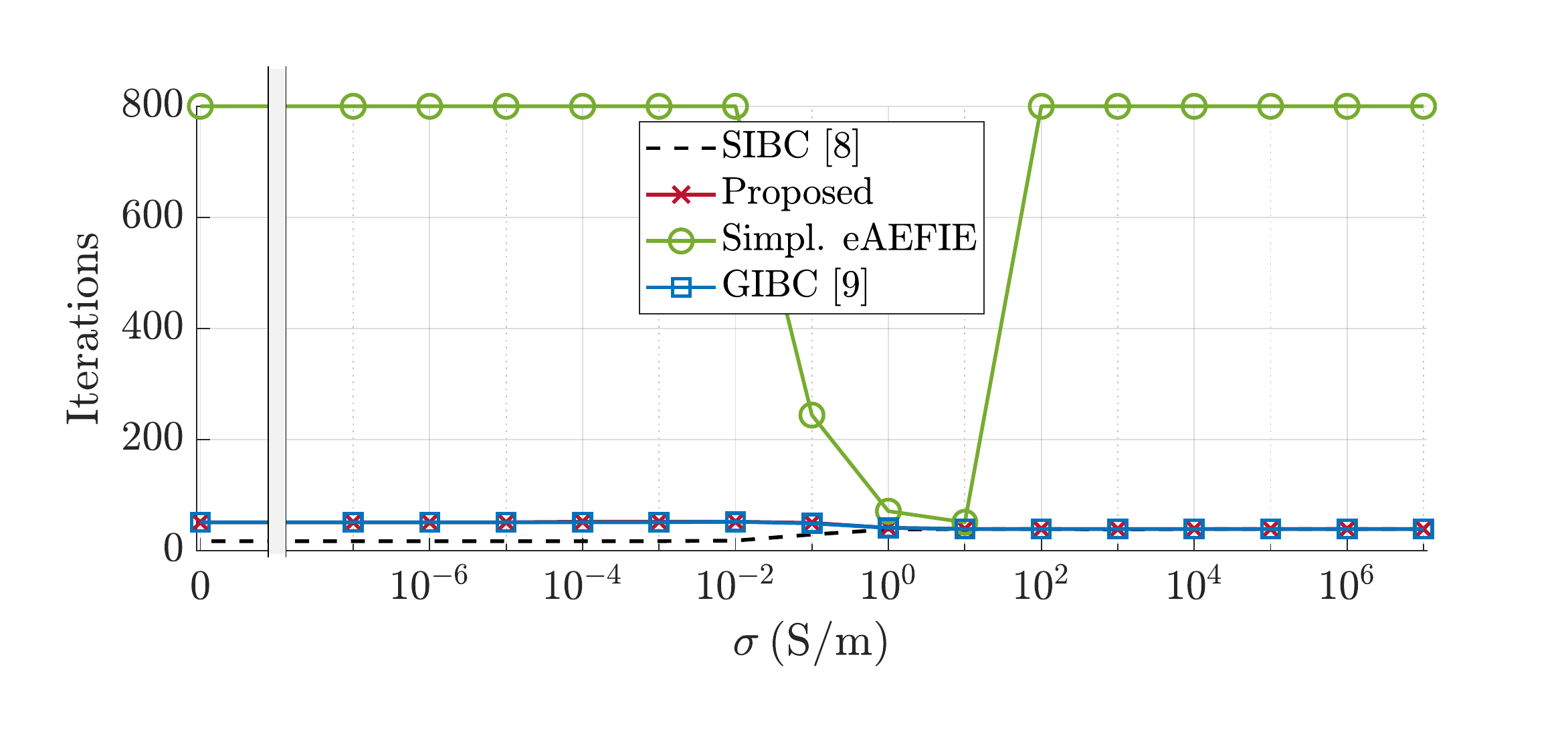}\label{fig:sphere_N_iter_sigma_1}
	}\\
	\subfloat[][]
	{
		\includegraphics[width=\linewidth,trim={0 12mm 0 14mm},clip=false]{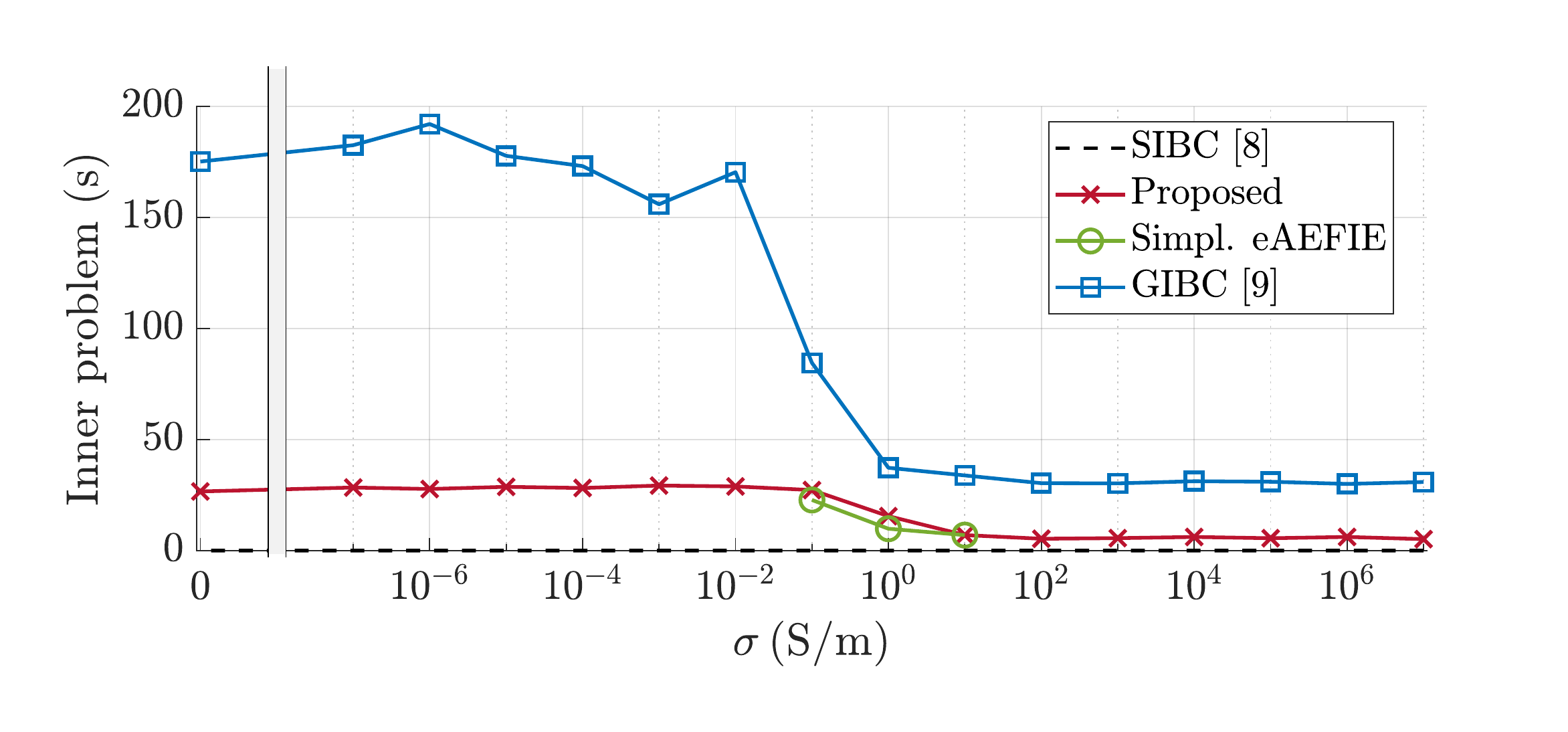}\label{fig:sphere_t_InnerModelGen_sigma_1}
	}\\
	\subfloat[][]
	{
		\includegraphics[width=\linewidth,trim={0 12mm 0 14mm},clip=false]{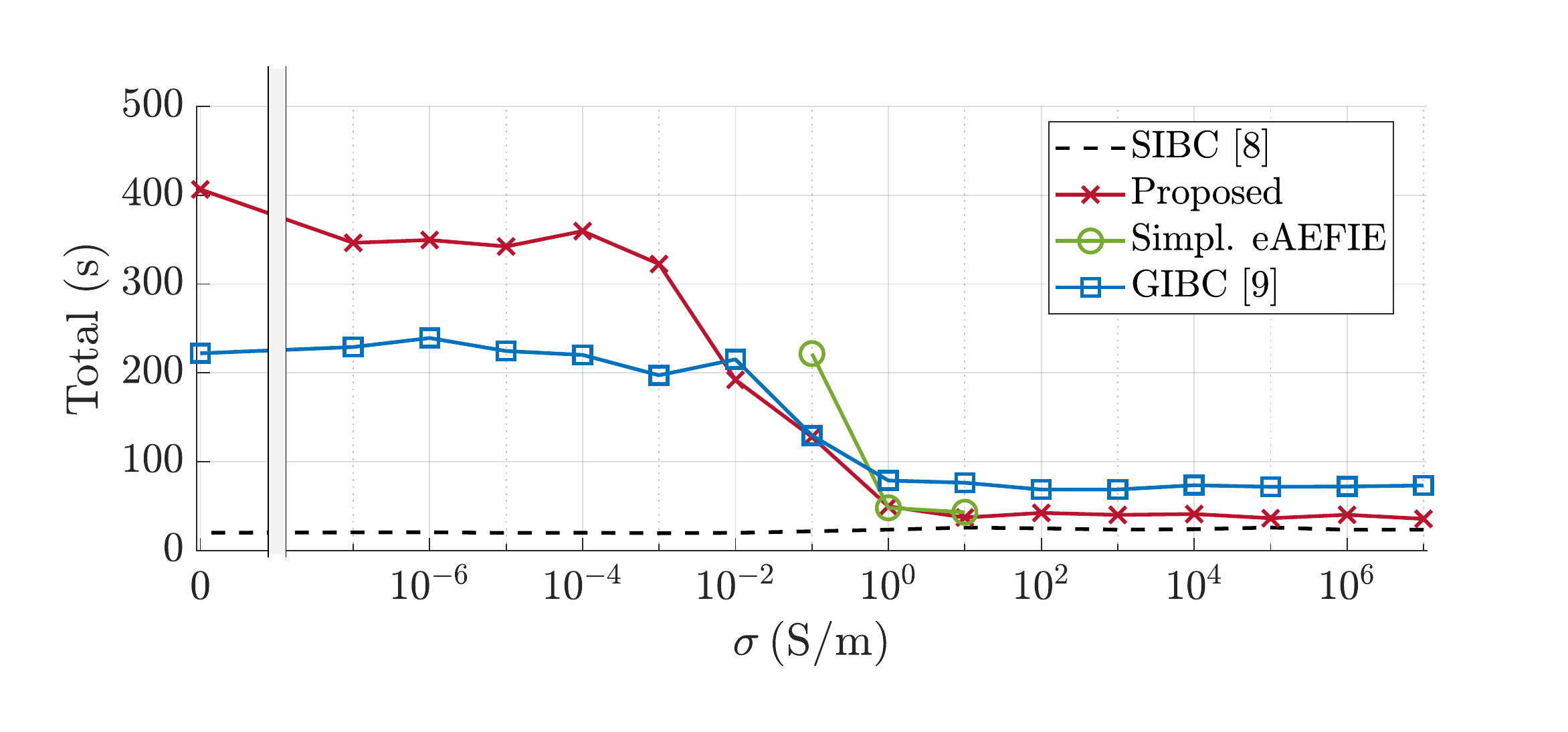}\label{fig:sphere_t_f_sigma_1}
	}
	\caption{Performance comparison for the sphere in \secref{sec:results:sphere_1}: (a) number of GMRES iterations required, (b) time taken in the interior problem, and (c) total time.}\label{fig:sphere_performance_sigma_1}
\end{figure}

\figref{fig:sphere_N_iter_sigma_1} shows the number of GMRES iterations required across a range of conductivities, for the proposed formulation compared to existing ones.
The maximum number of iterations allowed was capped at $800$.
It is clear that there is a significant advantage to the proposed and GIBC formulations, as compared to the simplified eAEFIE formulation.
Dual basis functions are required to improve the convergence of the eAEFIE, whereas the proposed method converges easily with RWG functions only.
In subsequent plots, results for the simplified eAEFIE are shown only for data points where the iterative solver converged.

The time taken in the interior problem for each of the formulations is shown in \figref{fig:sphere_t_InnerModelGen_sigma_1}.
As expected, the proposed method significantly reduces the interior problem time, particularly in the dielectric regime where there is no matrix sparsity induced by the skin effect.
The total simulation time for each formulation is shown in \figref{fig:sphere_t_f_sigma_1}.
In the dielectric regime, the proposed method is slower than the direct GIBC formulation because this is a relatively small problem, and directly generating the full GIBC operator is feasible and advantageous.
In the conductive regime, as skin effect develops, $\Lin$ becomes increasingly well conditioned due to the fast decay of the Green's functions.
This in turn speeds up the nested iterative solve in~\eqref{eq:nest2}, which in the limit $\sigma_i \rightarrow \infty$ approaches a direct solve (i.e.\,converges in one iteration).
Thus the savings in the interior problem have a more significant contribution at higher frequency, leading to better overall performance of the proposed method.
Due to the large number of iterations required by the simplified eAEFIE, this formulation performs well only for a narrow range of conductivities.

\subsection{Sphere: Scalability Study}\label{sec:results:scalability}
\begin{figure}[t]
	\centering
	\includegraphics[width=\linewidth]{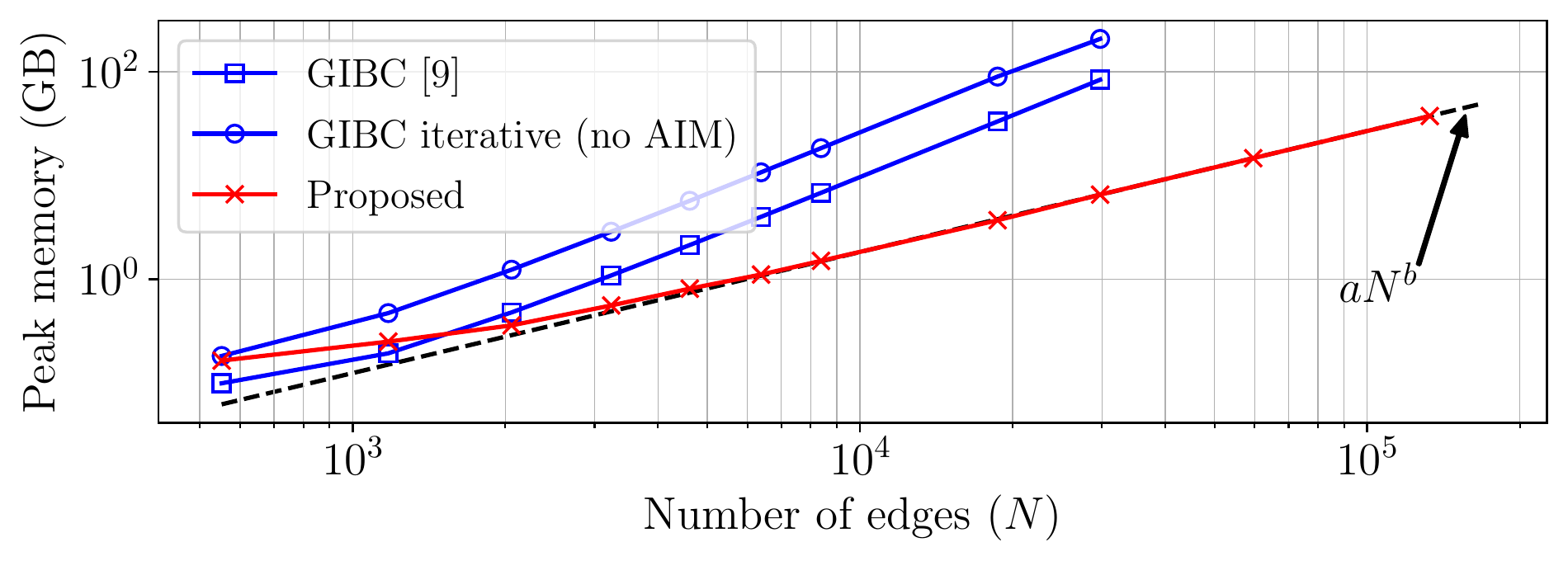}\\
	\includegraphics[width=\linewidth]{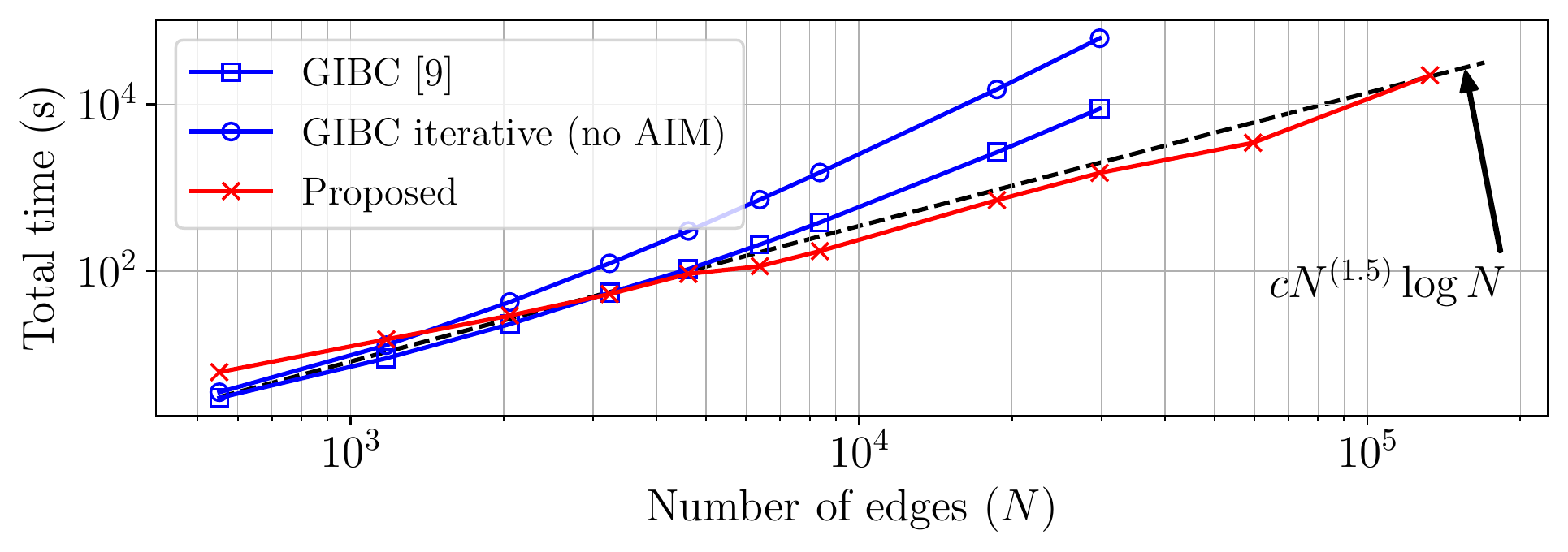}
	\caption{Scalability comparison for the sphere in \secref{sec:results:scalability}. Top panel: peak memory. Bottom panel: total time. Fit parameters are ${a = 3.93\times10^{-5}}\,$GB, ${b = 1.17}$, and ${c = 8.67\times10^{-5}}\,$s.}\label{fig:scalability}
\end{figure}

In this section, we study how the proposed method scales with the size of the problem compared to existing methods.
In general, the computational cost of the interior problem tends to be dominated by the number of mesh edges associated with the largest object in the structure.
Furthermore, the contribution of the interior problem to the overall cost depends on the size of each object in the structure.
Therefore, we consider for simplicity the case where the entire structure consists of a single object, and the goal is to demonstrate the need for the proposed method as the number of mesh edges increases.

We consider a sphere with diameter $1\,$m, meshed with a varying number of triangles.
The sphere has relative permittivity $2.5$, conductivity $0.1\,$S/m, and is excited with a plane wave at $100\,$MHz.
\figref{fig:scalability} shows the peak memory and total CPU time required as a function of the number of mesh edges, for the proposed method and for the original GIBC~\cite{GIBC}.
To demonstrate the need for the AIM in the interior problem, we also show the case where an iterative solver is used to compute the product $\matr{Z}_i\matr{H}_i^{(k)}$ at each iteration $k$, as in the proposed method, but without the use of AIM.

Both the original GIBC and the iterative GIBC without AIM have a memory cost of $\mathcal{O}(N^2)$ for~$N$ edges, due to the need to explicitly assemble~$\Lin$ and~$\Kin$.
Therefore, as shown in the top panel of \figref{fig:scalability}, both of these methods are only able to handle approximately $20{,}000$ edges at most, within the available $256\,$GB of memory.
The proposed method has an expected memory cost of approximately $\mathcal{O}(N)$ due to the use of the AIM.
Up to approximately $1{,}000$ edges, the actual memory required by the proposed method is comparable to the other two methods.
Beyond that, the proposed method requires significantly less memory than the original and iterative GIBC methods, and scales easily to over $100{,}000$ edges.
A least squares fit is also shown in the top panel of \figref{fig:scalability}, which indicates that the memory cost of the proposed method scales as $\mathcal{O}(N^{1.17})$, which is slightly larger than the expected $\mathcal{O}(N)$, likely due to implementation-related overhead.
The iterative GIBC without AIM requires more memory than the original GIBC because it requires storing both~$\Lin$ and~$\Kin$ separately.

The original GIBC has a time complexity of $\mathcal{O}(N^3)$ due to the factorization of $\Lin$.
The iterative GIBC without the AIM has time complexity of $\mathcal{O}(N^2)$, due to the need to assemble and multiply the dense matrices $\Lin$ and $\Kin$.
The expected time cost associated with the proposed method is approximately $\mathcal{O}(N^{1.5}\log N)$.
The relative efficiency of the proposed method can be seen in the bottom panel of \figref{fig:scalability}, which confirms the $\mathcal{O}(N^{1.5}\log N)$ scaling of the proposed method via a least squares fit.
Up to approximately $2{,}000$ edges, all three methods have a comparable cost.
As the number of edges exceeds that amount, the proposed method becomes significantly faster than the original GIBC and the iterative GIBC without AIM.

\subsection{Multiscale Split Ring Resonator Array}\label{sec:results:SRR_5}

\begin{figure}[t]
	\centering
	\includegraphics[width=\linewidth]{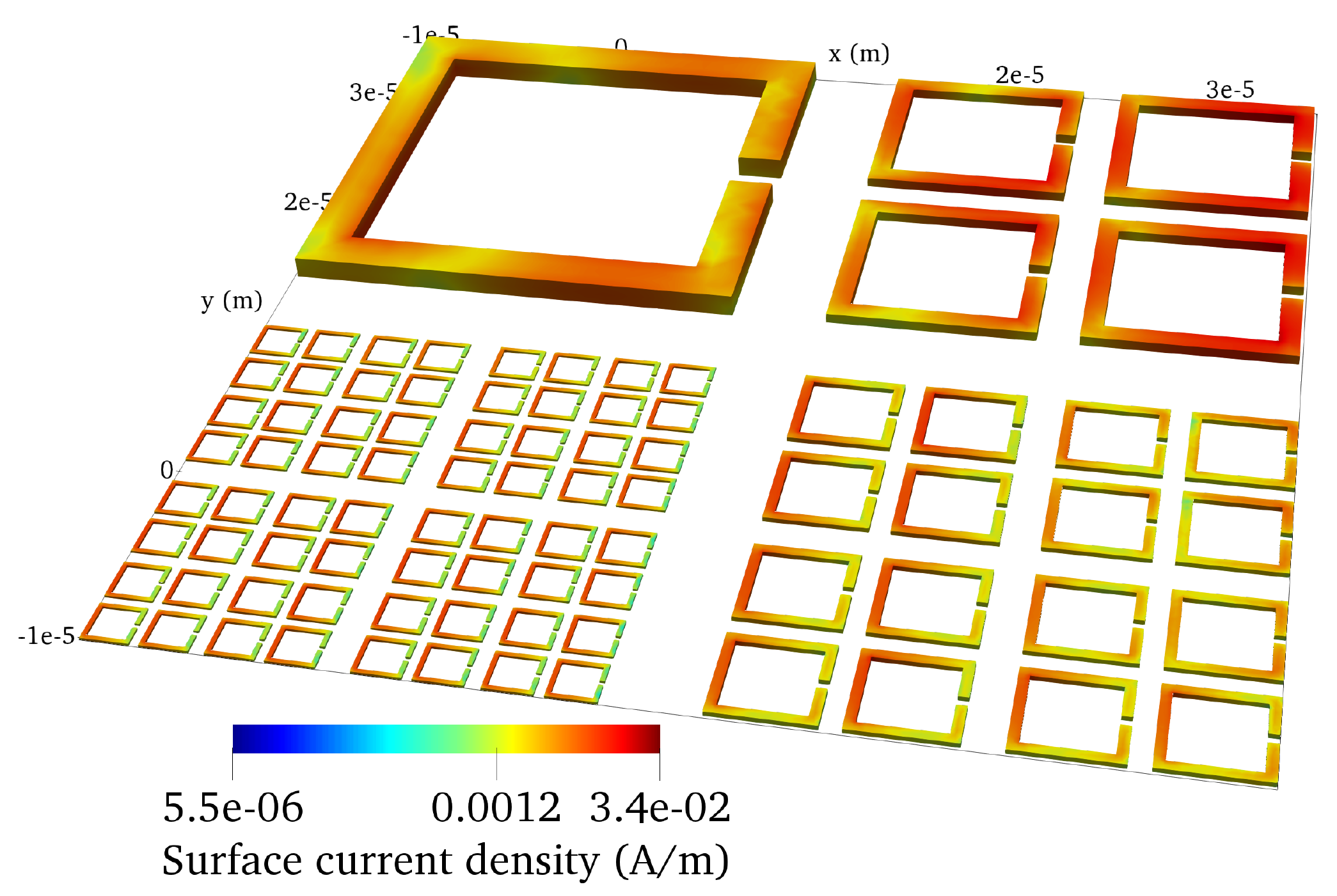}
	\caption{Geometry of the multiscale SRR array in \secref{sec:results:SRR_5}, with the electric surface current density at $10\,$THz.}\label{fig:J_SRR_5}
\end{figure}

\begin{figure}[t]
  \centering
  \includegraphics[width=\linewidth,trim={0 12mm 0 14mm},clip=false]{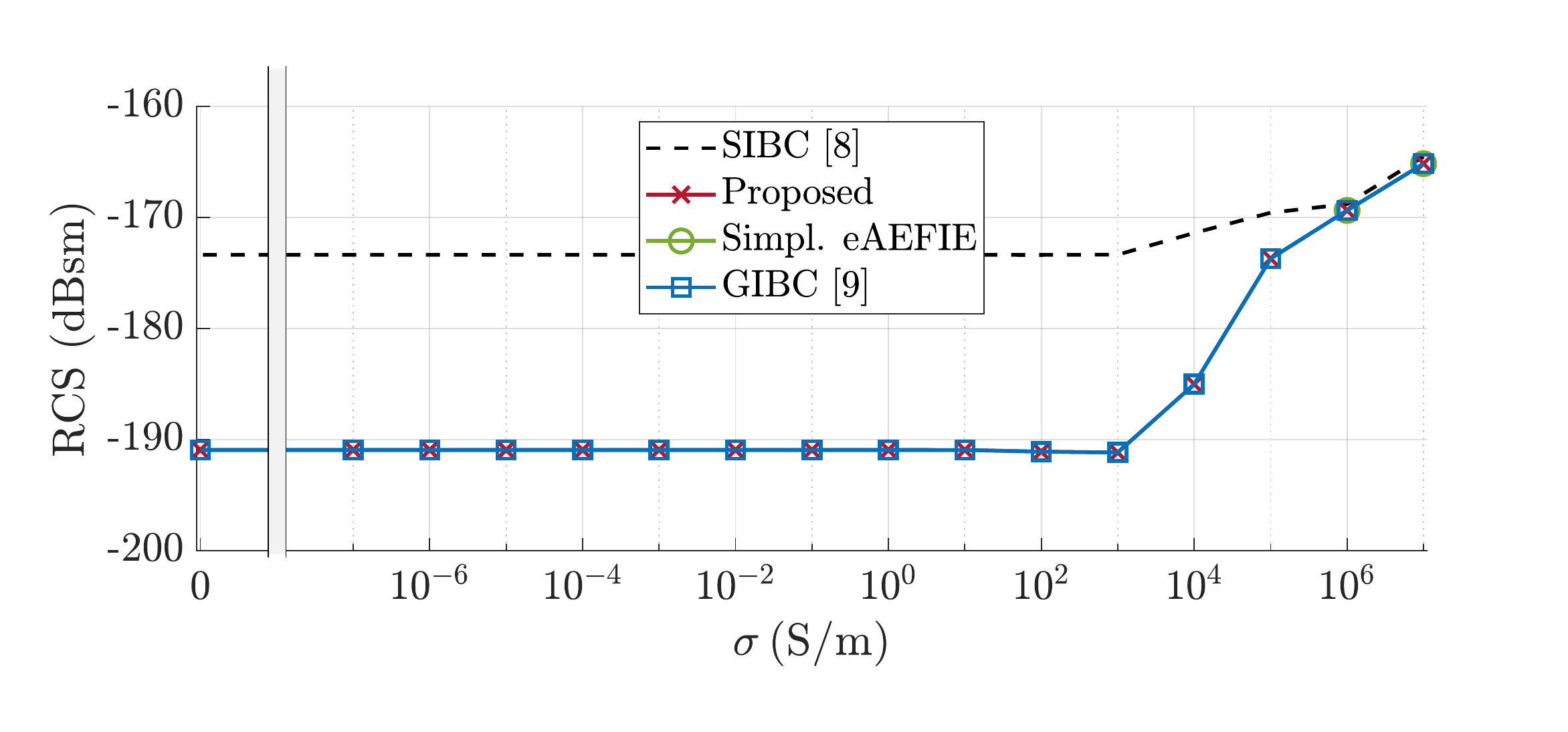}
  \caption{Monostatic RCS of the SRR array in \secref{sec:results:SRR_5} as a function of conductivity, at $10\,$THz.}\label{fig:SRR_mRCS_sigma_5}
\end{figure}

Next, we consider a multiscale arrangement of split ring resonators (SRRs), as shown in \figref{fig:J_SRR_5}.
The structure consists of 85 SRRs with side length and gap width ranging from $2\,\mu\mathrm{m}$ to $20\,\mu\mathrm{m}$, and thickness ranging from $0.1\,\mu\mathrm{m}$ to $1\,\mu\mathrm{m}$.
The relative permittivity of each SRR is set to $11$, and the structure is meshed with $75{,}310$ triangles and $112{,}965$ edges.

Both conductive and dielectric SRR arrays in this size range are of practical importance in metasurface and metamaterial applications~\cite{SRRmetaCond,SRRmetaDiel}, making it a useful test case to assess the versatility of the proposed method.
Therefore, we consider a range of conductivity from $0$ to $10^7\,$S/m, which spans both dielectric and near-PEC regimes.

\begin{figure}[t]
	\centering
	\subfloat[][]
	{
		\includegraphics[width=\linewidth,trim={0 12mm 0 14mm},clip=false]{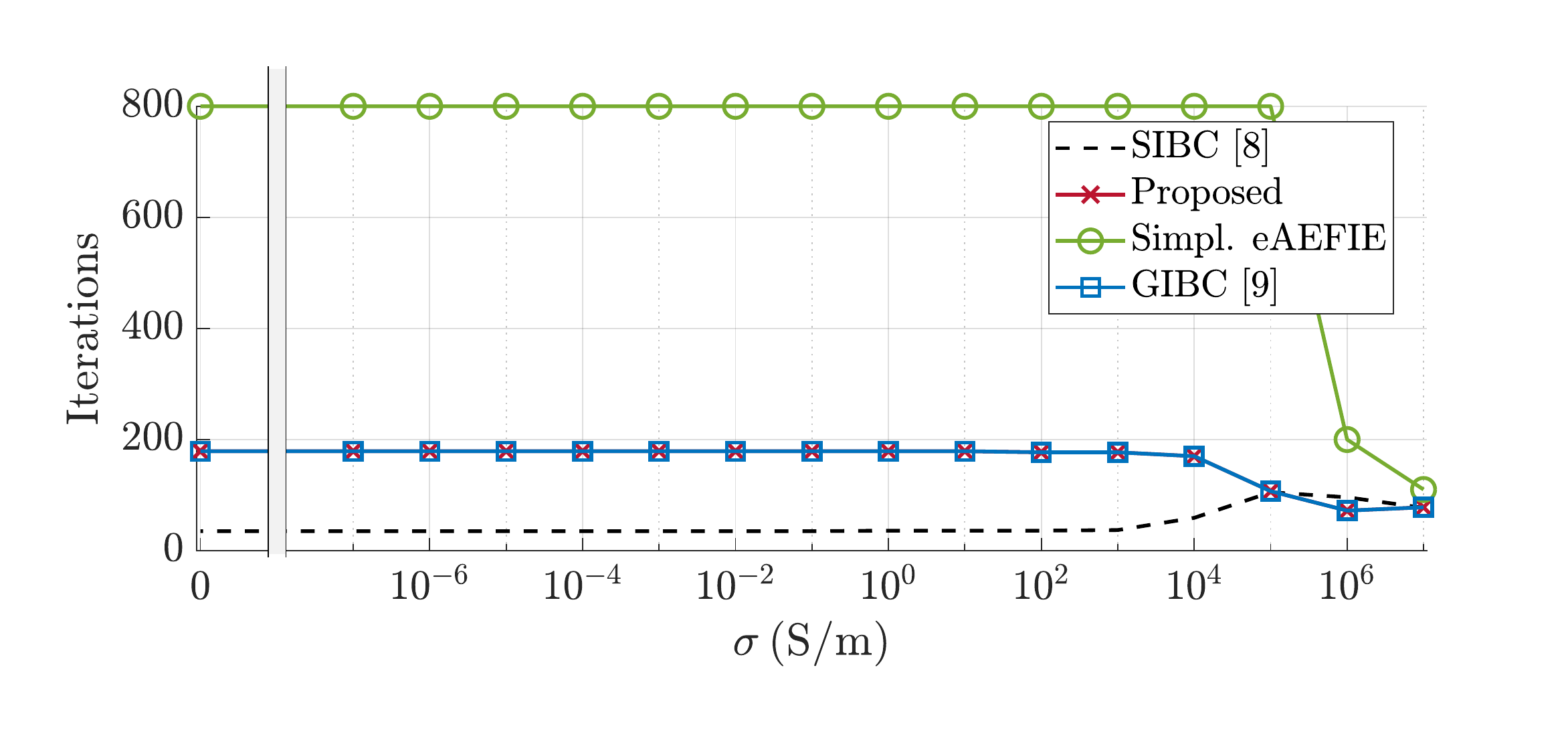}\label{fig:SRR_N_iter_sigma_5}
	}\\
	\subfloat[][]
	{
		\includegraphics[width=\linewidth,trim={0 12mm 0 14mm},clip=false]{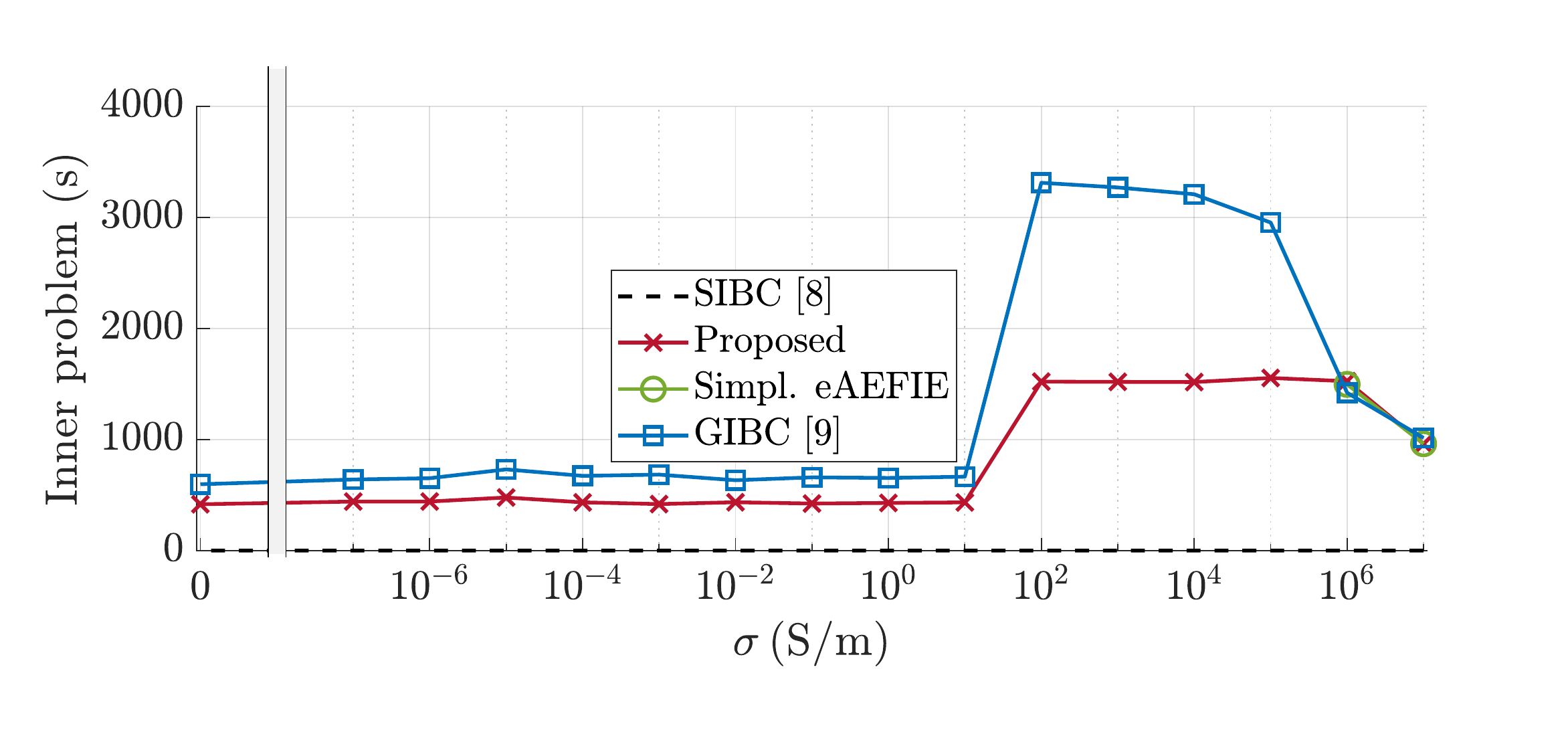}\label{fig:SRR_t_InnerModelGen_sigma_5}
	}\\
	\subfloat[][]
	{
		\includegraphics[width=\linewidth,trim={0 12mm 0 14mm},clip=false]{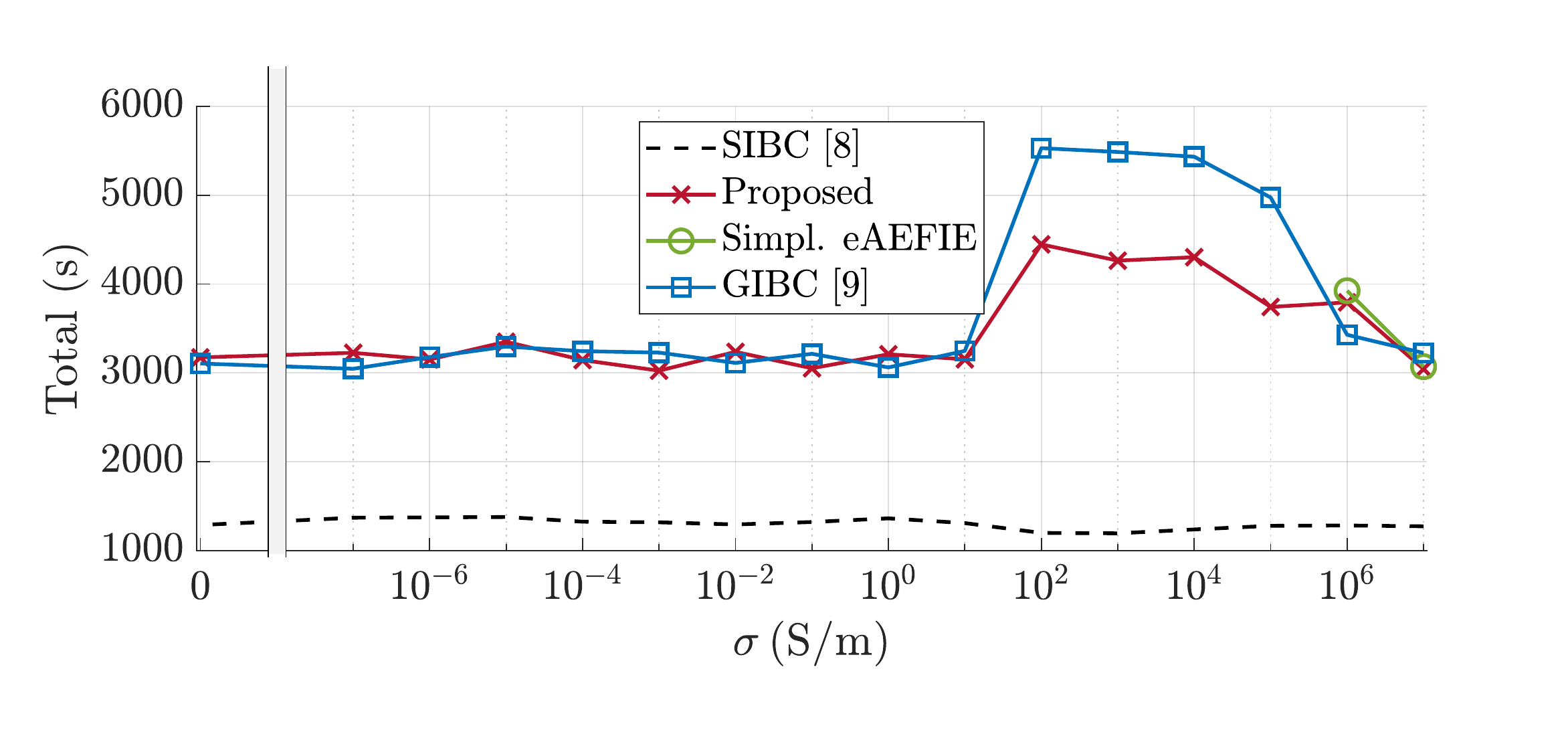}\label{fig:SRR_t_f_sigma_5}
	}
	\caption{Performance comparison for the multiscale SRR array in \secref{sec:results:SRR_5}: (a) number of GMRES iterations required, (b) time taken in the interior problem, and (c) total time.}\label{fig:SRR_performance_sigma_5}
\end{figure}

The structure is excited by a plane wave at $10\,$THz, with the electric field polarized along the $y$ axis.
\figref{fig:J_SRR_5} shows the surface current distribution for ${\sigma = 10}\,$S/m, and \figref{fig:SRR_mRCS_sigma_5} shows that the monostatic RCS computed with the proposed method is in good agreement with existing methods.
The SIBC approximation is inaccurate except for large conductivities, as expected.
\figref{fig:SRR_N_iter_sigma_5} demonstrates the that the proposed method is well conditioned across the entire range of conductivities considered, whereas the simplified eAEFIE only converges for large values of conductivity.
In subsequent plots, results for the simplified eAEFIE are shown only for data points where the iterative solver converged.

\figref{fig:SRR_t_InnerModelGen_sigma_5} shows the reduction in interior problem time of the proposed method, compared to the original GIBC formulation.
The proposed method is significantly faster only for conductivities between $10^2\,$S/m and $10^5\,$S/m, and comparable to the original GIBC for other conductivities, because the objects in this example are small. The mesh for the largest split ring has only $1{,}272$ edges, so the size of the interior problem matrices $\Lin$ and $\Kin$ is at most ${1{,}272 \times 1{,}272}$.
Therefore, the direct factorization of~$\Lin$ in~\eqref{eq:GIBCdef} is relatively easy on a modern desktop computer, which makes the original GIBC computationally feasible.
The primary time savings in the proposed method are due to the fact that only near-region entries of $\Lin$ and $\Kin$ need to be computed.
For small values of conductivity, these savings are less noticeable because the proposed approach also requires the AIM precorrection step, unlike the original GIBC.
For larger values of conductivity, the savings are more apparent because the cost of computing the entries of $\Lin$ and $\Kin$ is dominant, since expensive numerical integration techniques~\cite{GIBC,eaefie02} are required to capture the faster oscillations of the associated Green's function.
This is evidenced by the increase in inner problem time visible in \figref{fig:SRR_t_InnerModelGen_sigma_5} near ${\sigma_i = 100}\,$S/m, when the advanced numerical integration technique proposed previously~\cite{GIBC} is employed.
As conductivity is increased beyond $10^5\,$S/m, skin effect develops, leading to sparser matrices and therefore a decrease in inner problem time for $\sigma \gtrsim 10^5\,$S/m.
Finally, \figref{fig:SRR_t_f_sigma_5} shows the overall superiority of the proposed method in terms of total simulation time, when the entire range of conductivities is taken into consideration.

\subsection{Integrated Circuit (IC) Package}\label{sec:results:pckg}

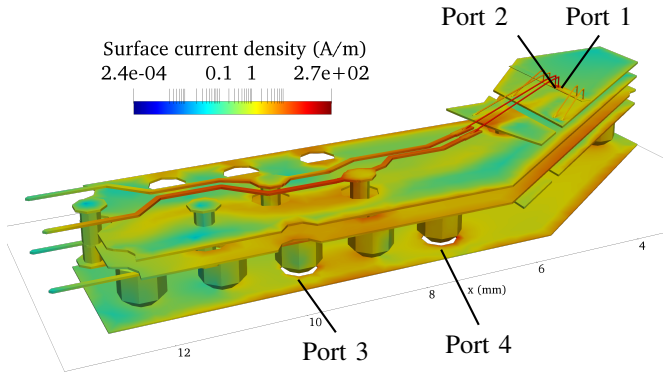
\begin{figure}[t]
	\centering
	\begin{tikzpicture}
		\node[] at (0,0) {\includegraphics[width=\linewidth]{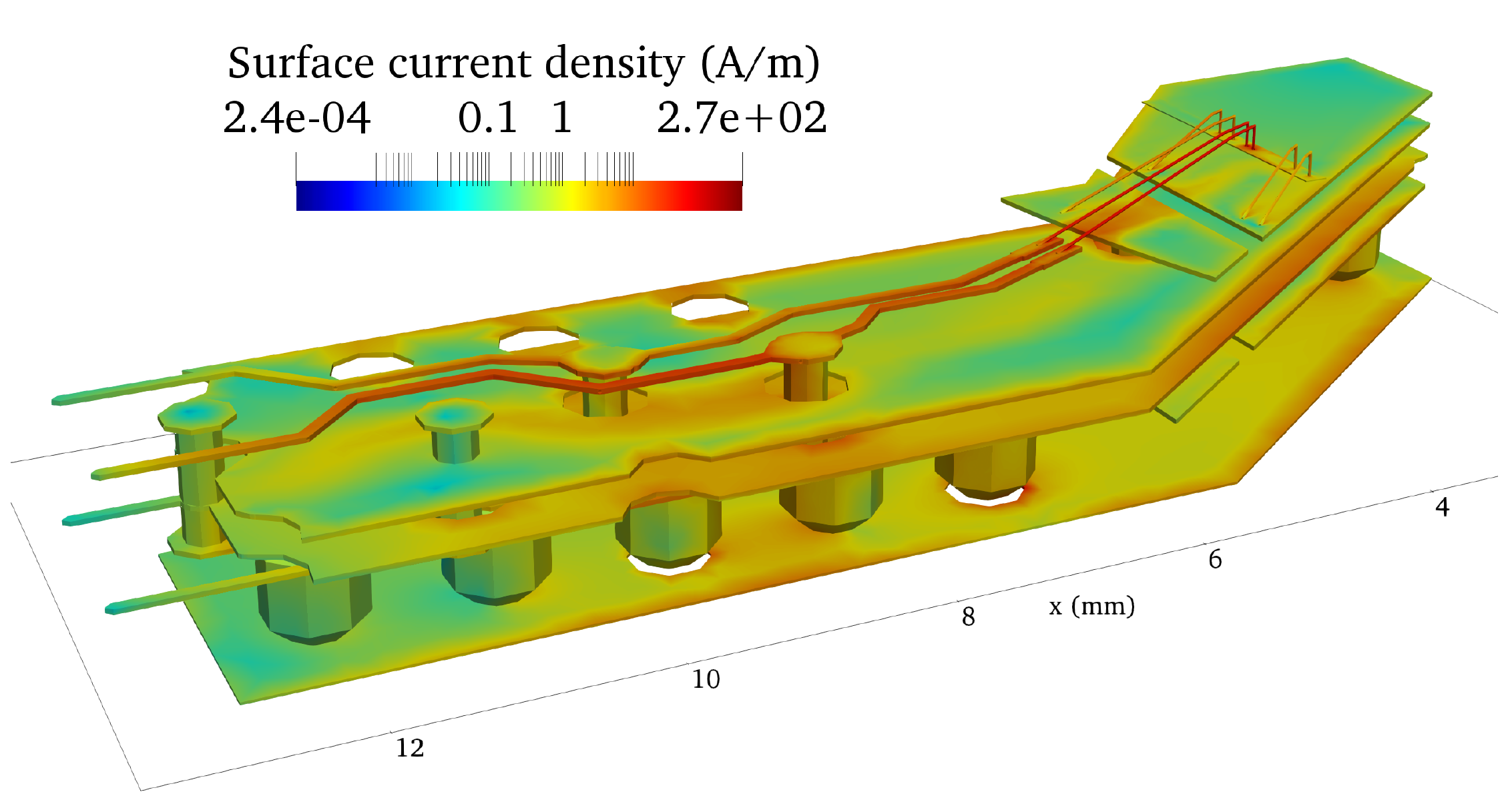}};
		\draw[thick] (3.0,1.5)--+(0.5,0.7)node[above]{Port 1};
		\draw[thick] (2.92,1.5)--+(-0.5,0.7)node[above left]{Port 2};
		\draw[thick] (-0.5,-1.05)--+(0.5,-0.7)node[below]{Port 3};
		\draw[thick] (1.4,-0.65)--+(0.5,-1.0)node[below]{Port 4};
	\end{tikzpicture}
	\caption{Geometry of the IC package in \secref{sec:results:pckg}, with the electric surface current density at $10\,$GHz.}\label{fig:J_pckg}
\end{figure}

\begin{figure}[t]
	\centering
	\includegraphics[width=\linewidth]{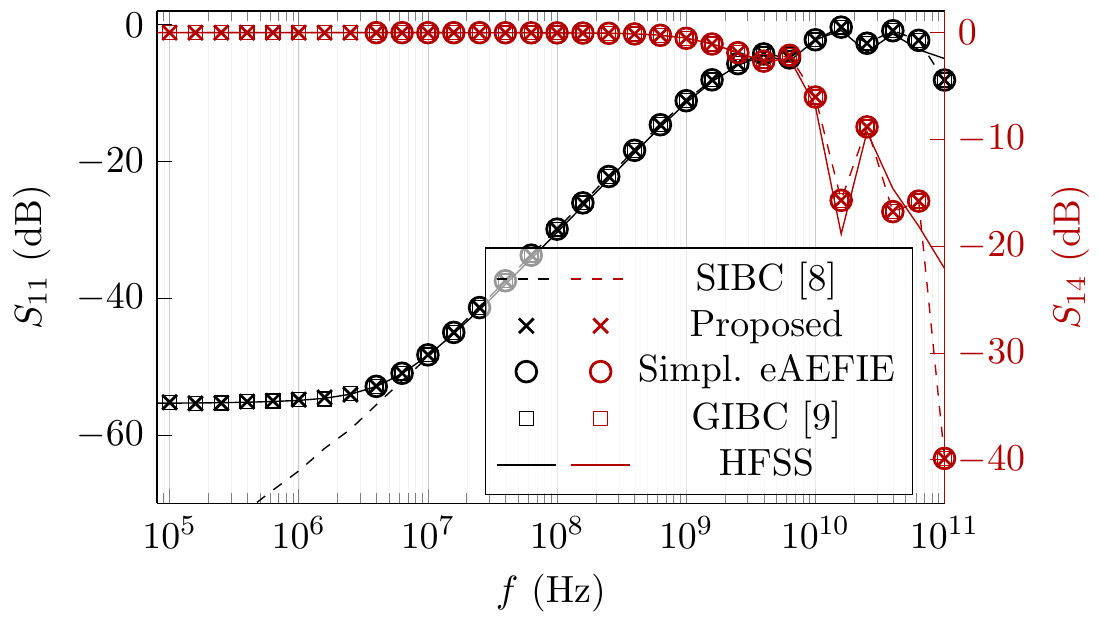}
	\caption{Scattering parameters for the IC package in \secref{sec:results:pckg}.}\label{fig:pckg_S}
\end{figure}

We consider next a part of an IC package provided through a commercial finite element tool (Ansys HFSS 2019).
The four-port copper structure is meshed with 31,718 triangles and 47,577 edges.
A Th\'evenin equivalent circuit excitation is used at each port to extract scattering ($S$) parameters~\cite{gope,CPMT2019arxiv}.

The geometry and electric surface current density at $10\,$GHz are shown in \figref{fig:J_pckg}.
The $S$ parameters are shown in \figref{fig:pckg_S} for a broad frequency band from $10\,$kHz to $100\,$GHz.
The proposed method is in excellent agreement with the direct GIBC formulation.
At frequencies below approximately $10\,$MHz, when the skin effect has not fully developed, the SIBC approximation is inaccurate.
At high frequencies, above approximately $10\,$GHz, the finite element mesh of HFSS is not fine enough to model the narrow skin depth, lower than $0.7\,\mu\mathrm{m}$.
Further mesh refinement led to physical memory requirements that were beyond the available 256\,GB.
The proposed method agrees well with the SIBC approximation at high frequencies, and with HFSS at low and intermediate frequencies.

\begin{figure}[t]
	\centering
	\subfloat[][]
	{
		\includegraphics[width=\linewidth,trim={0 12mm 0 14mm},clip=false]{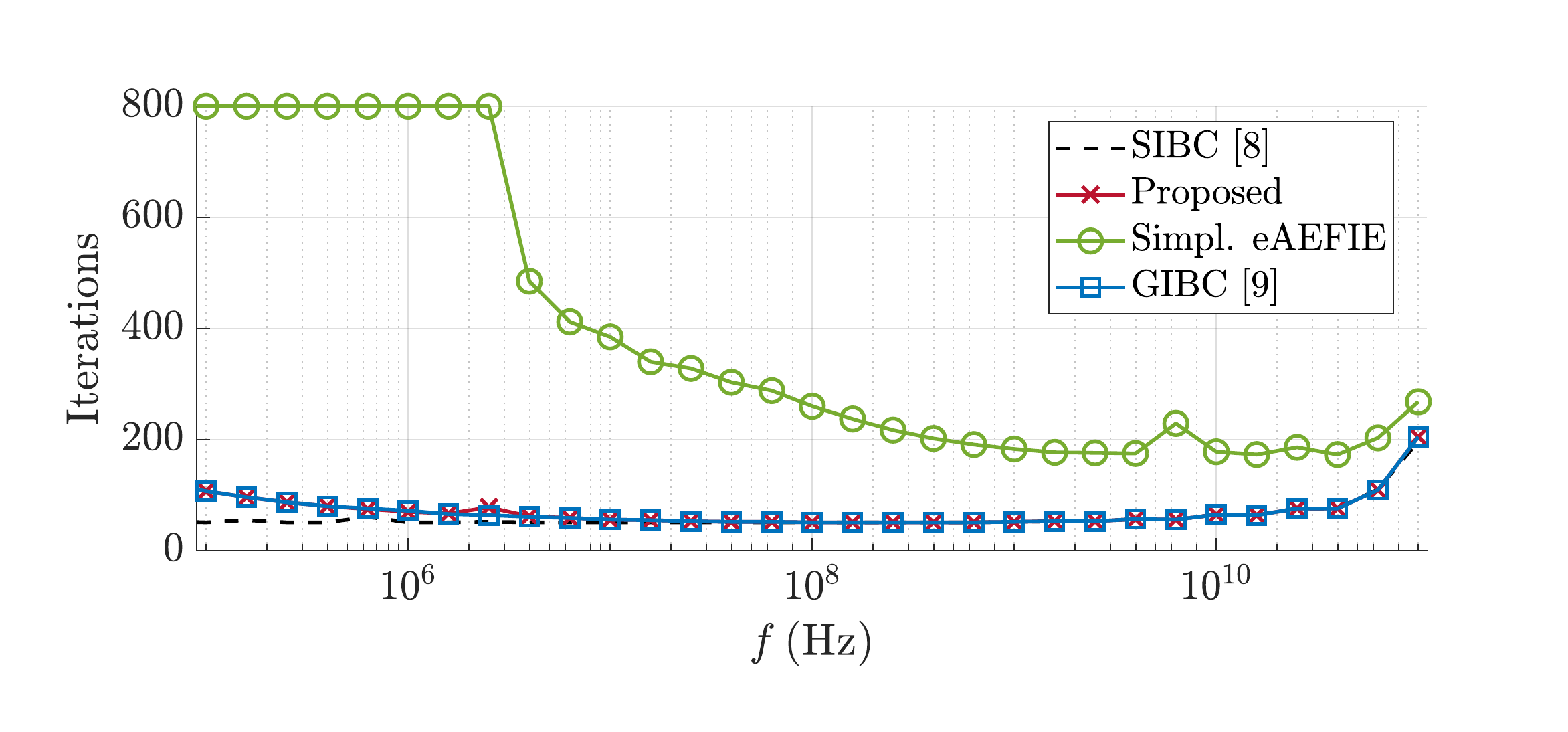}\label{fig:pckg_N_iter}
	}\\
	\subfloat[][]
	{
		\includegraphics[width=\linewidth,trim={0 12mm 0 14mm},clip=false]{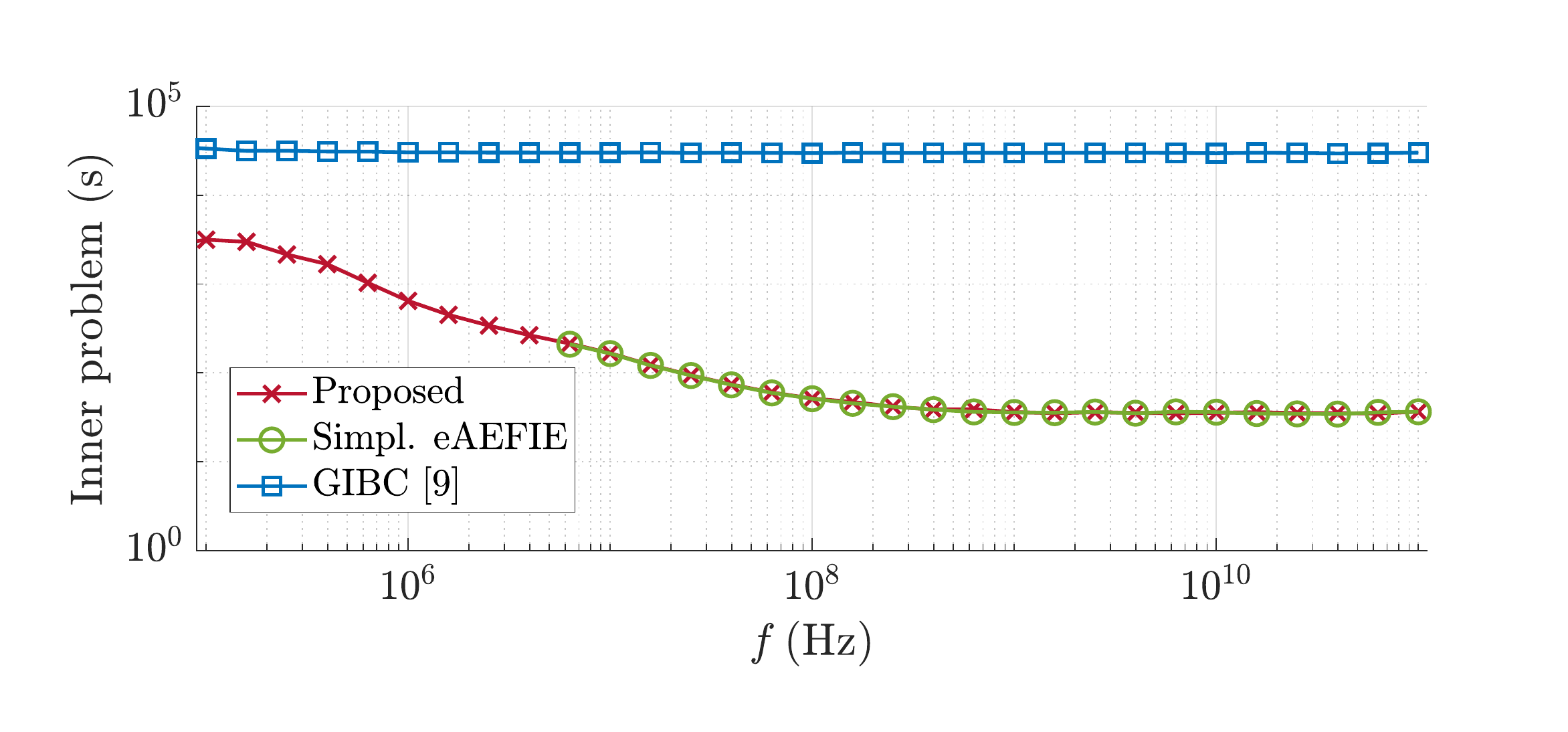}\label{fig:pckg_t_InnerModelGen}
	}\\
	\subfloat[][]
	{
		\includegraphics[width=\linewidth,trim={0 12mm 0 14mm},clip=false]{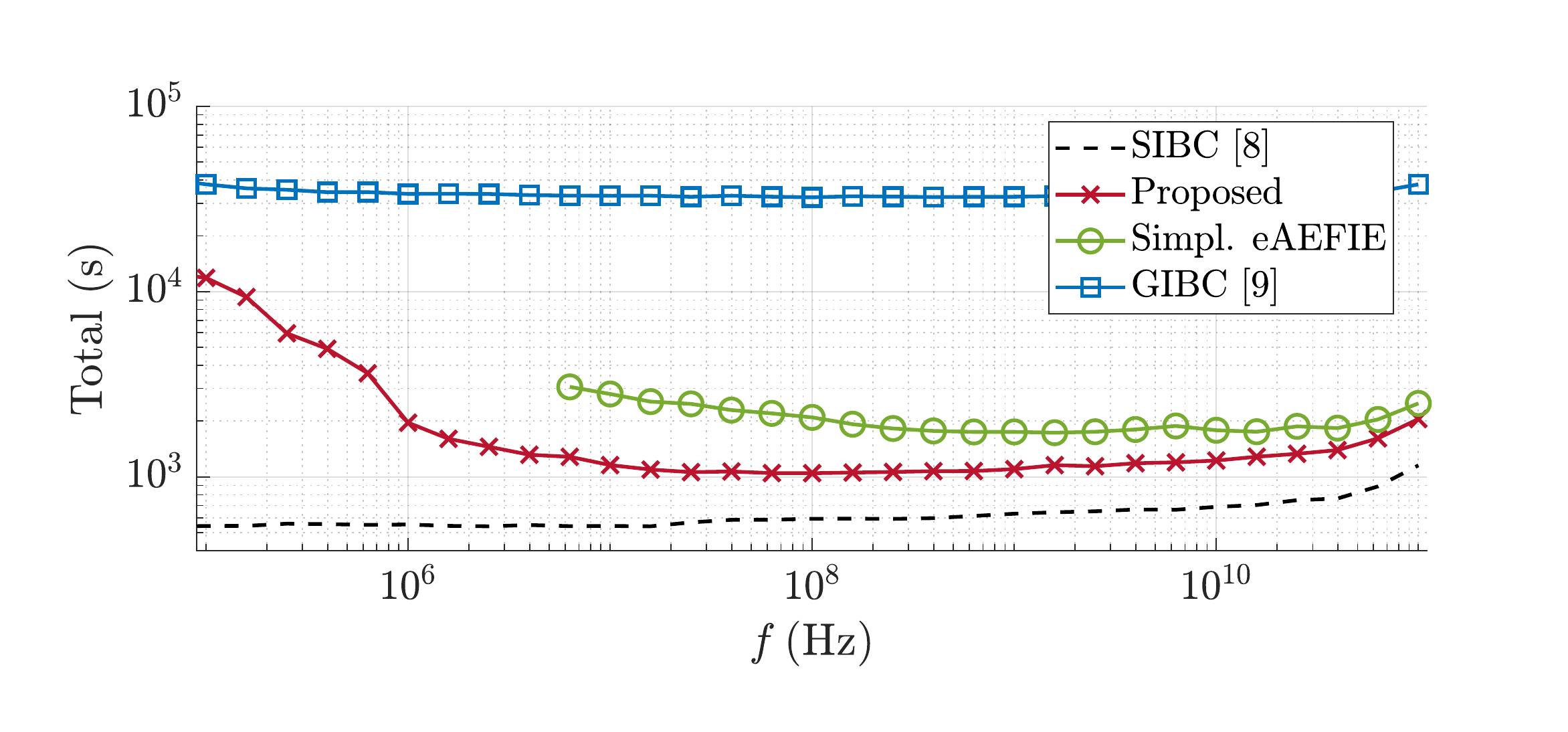}\label{fig:pckg_t_f}
	}
	\caption{Performance comparison for the IC in \secref{sec:results:pckg}: (a) number of GMRES iterations for the exterior problem, (b) time taken in the interior problem, and (c) total time.}\label{fig:pckg_performance}
\end{figure}

\figref{fig:pckg_N_iter} shows the number of GMRES iterations required for convergence of the exterior problem.
The proposed method is very well conditioned over the entire frequency range, while the simplified eAEFIE consistently requires a significantly larger number of iterations to converge.
\figref{fig:pckg_t_InnerModelGen} shows the time taken to generate the interior problem operators.
The proposed method is more efficient than the direct GIBC formulation by a factor ranging from~$10\times$ to~$800\times$.
The IC package consists of several large objects, some having meshes with over~$10{,}000$ edges,
Hence the interior problem time in the direct GIBC formulation is dominated by the assembly of the dense operators~$\matr{Z}_i$.
At sufficiently high frequencies, when the interior problem matrices become sparse, it is possible to make the direct GIBC method more efficient by storing the sparse matrices~$\Lin$ and~$\Kin$ separately, rather than explicitly assembling the dense operators~$\matr{Z}_i$ by factorizing~$\Lin$.
However, this would require setting a heuristic threshold to switch between two different sets of codes for generating the operator and computing the matrix-vector product.
The proposed method achives this in a seamless way, as evidenced by the decrease in inner problem times with frequency.

\begin{figure}[t]
\centering
	\includegraphics[width=\linewidth,trim={0 8mm 0 10mm},clip=true]{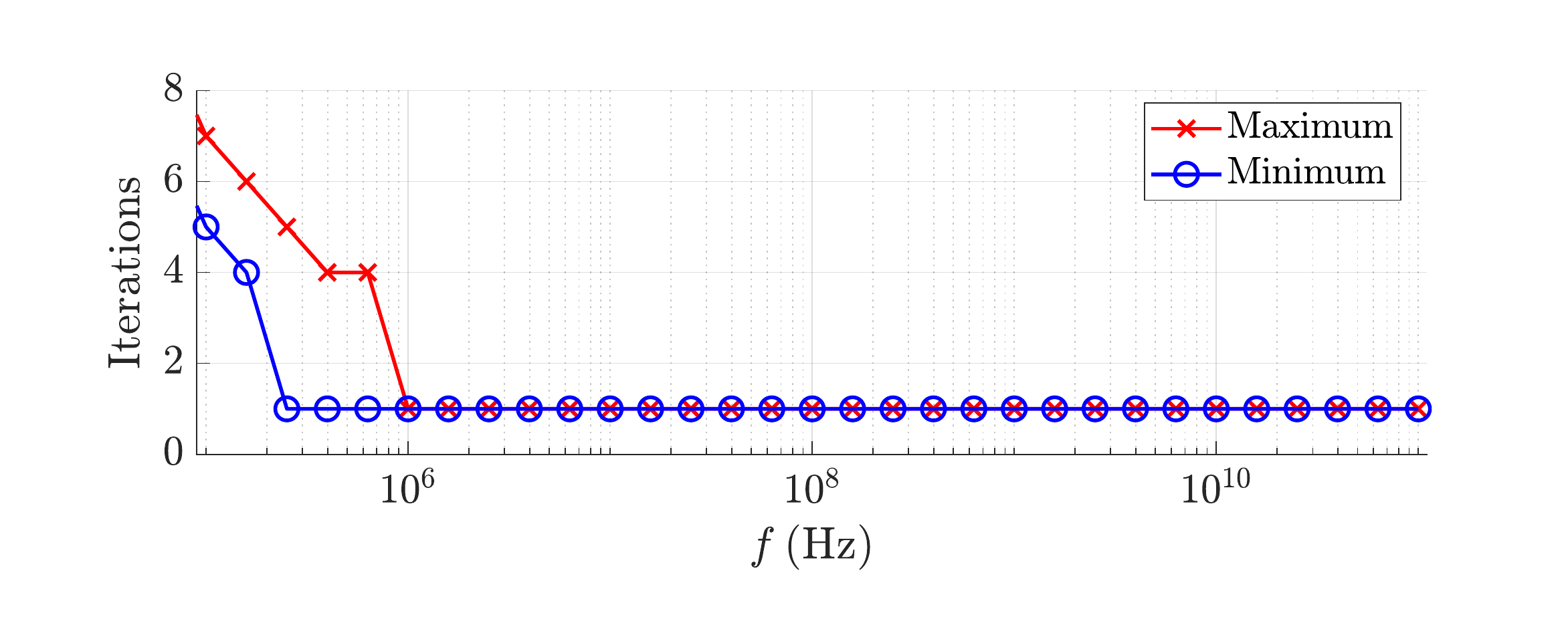}
	\caption{Maximum and minimum number of GMRES iterations required for the interior problem in the proposed method, for the IC package in \secref{sec:results:pckg}.}\label{fig:pckg_N_iter_inner}
\end{figure}

\figref{fig:pckg_t_f} reports the total simulation time per frequency for each method, and indicates the overall superior performance of the proposed method over the entire frequency range.
Finally, \figref{fig:pckg_N_iter_inner} shows the maximum and minimum number of nested iterations required by the accelerated surface impedance operator in the proposed method.
The two curves correspond to the nested interior problems of two different objects of the structure.
Convergence is achieved within $10$ iterations at most, and improves at higher frequency as $\Lin$ becomes increasingly diagonally dominant.
This confirms the effectiveness of the proposed near-region preconditioner~\eqref{eq:nest2pc} for the interior problem.

\subsection{On-Chip Interconnect Network}\label{sec:results:interposer}

\begin{figure}[t]
  \centering
  \includegraphics[width=.9\linewidth]{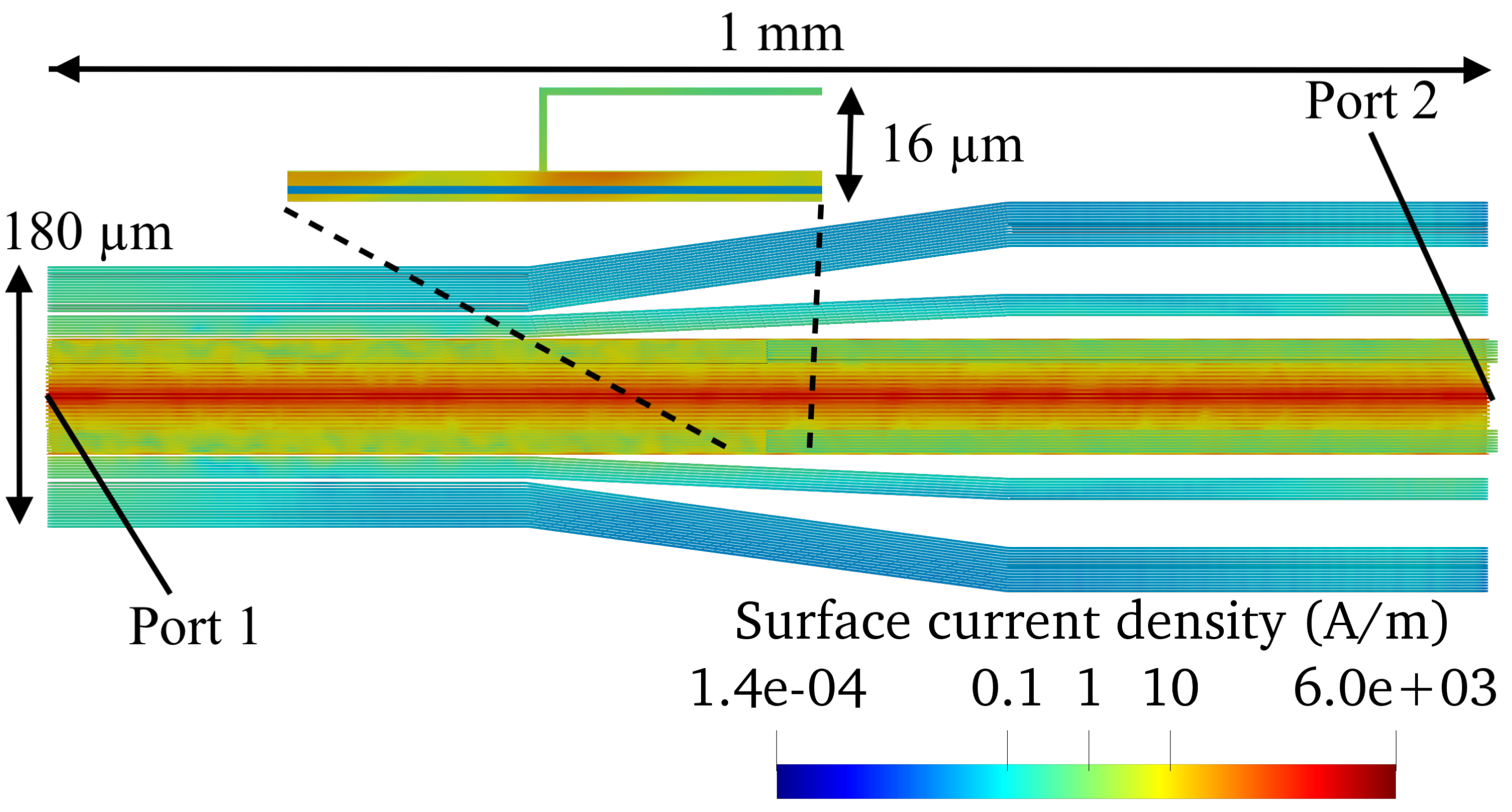}
  \caption{Geometry of the interconnect network in \secref{sec:results:interposer}, with the electric surface current density at $1\,$GHz.}\label{fig:J_interposer}
\end{figure}

\begin{figure}[t]
	\centering
	\includegraphics[width=\linewidth]{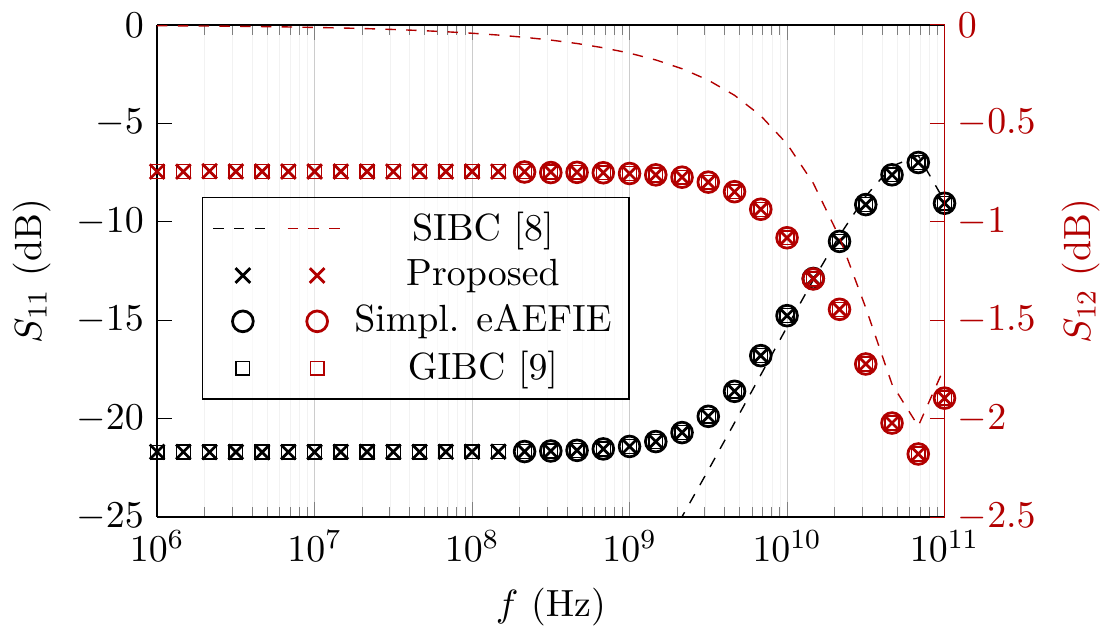}
	\caption{Scattering parameters for the interconnect network in \secref{sec:results:interposer}.}\label{fig:interposer_S}
\end{figure}

Finally, we consider an on-chip interconnect network consisting of 80 copper signal lines and a ground plane, in free space.
Similar networks are encountered in interposers for applications such as high-bandwidth memory~\cite{HBM01}.
The network is meshed with 194,464 triangles and 291,696 edges.
The geometry of the network and electric surface current density at $1\,$GHz are shown in \figref{fig:J_interposer}.

\figref{fig:interposer_S} shows the $S$ parameters over a wide frequency band.
The proposed method is in good agreement with the direct GIBC method.
Again, as expected, the SIBC approximation is only accurate at high frequencies.
HFSS is inaccurate at high frequencies when the skin depth is extremely small.
The simplified eAEFIE does not yield accurate $S$ parameters at low frequencies because it does not converge within $800$ iterations.

\begin{figure}[t]
	\centering
	\subfloat[][]
	{
		\includegraphics[width=\linewidth,trim={0 12mm 0 14mm},clip=false]{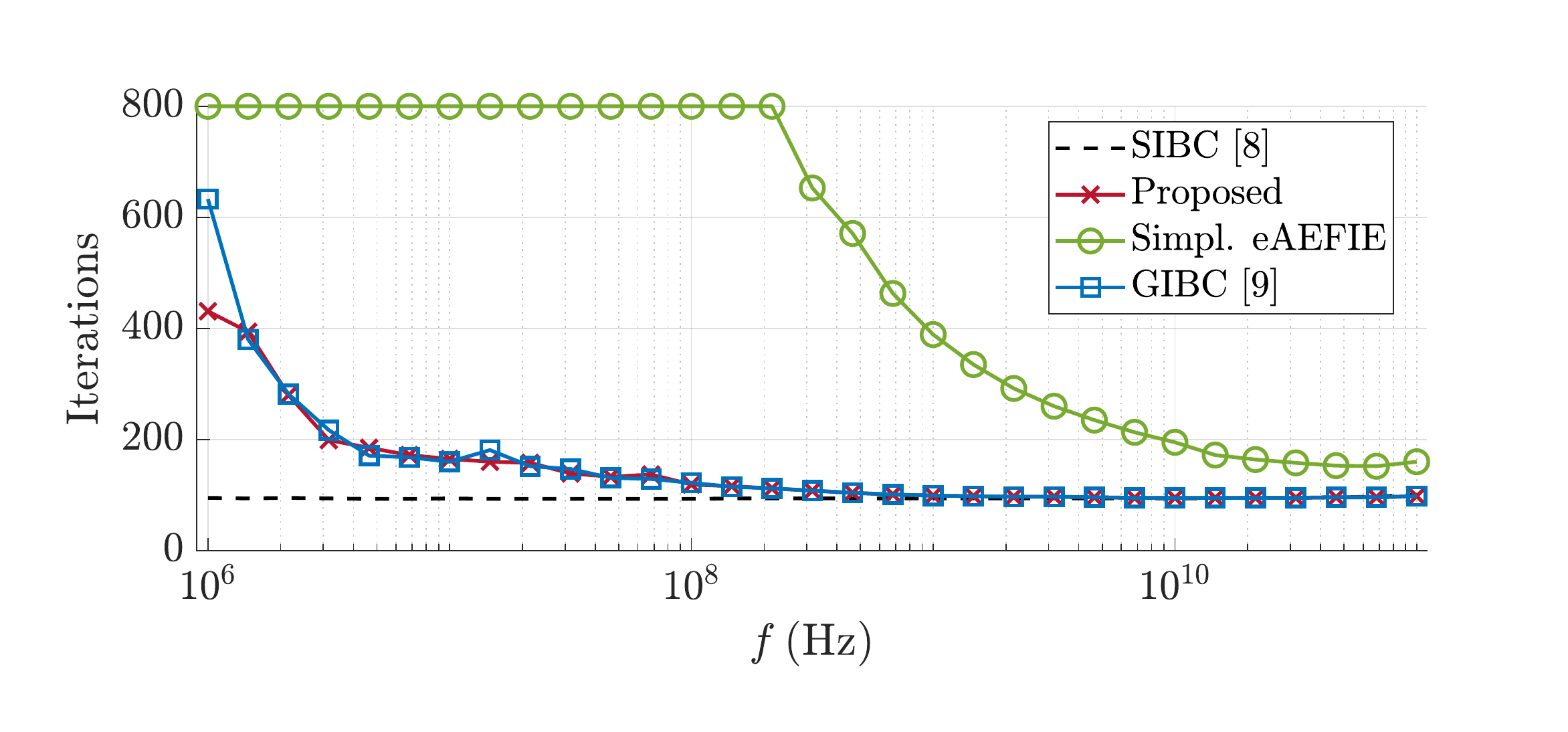}\label{fig:interposer_N_iter}
	}\\
	\subfloat[][]
	{
		\includegraphics[width=\linewidth,trim={0 12mm 0 14mm},clip=false]{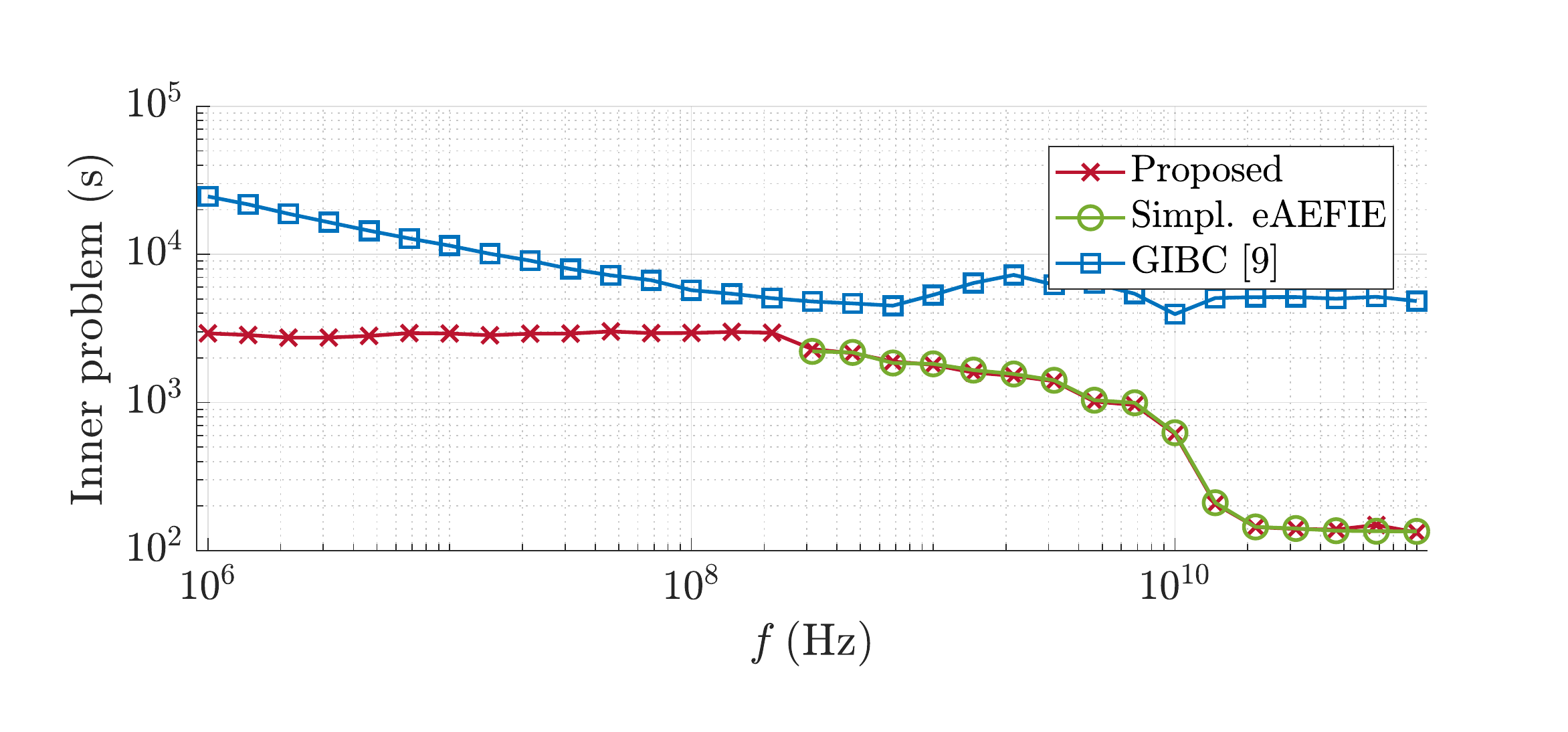}\label{fig:interposer_t_InnerModelGen}
	}\\
	\subfloat[][]
	{
		\includegraphics[width=\linewidth,trim={0 12mm 0 14mm},clip=false]{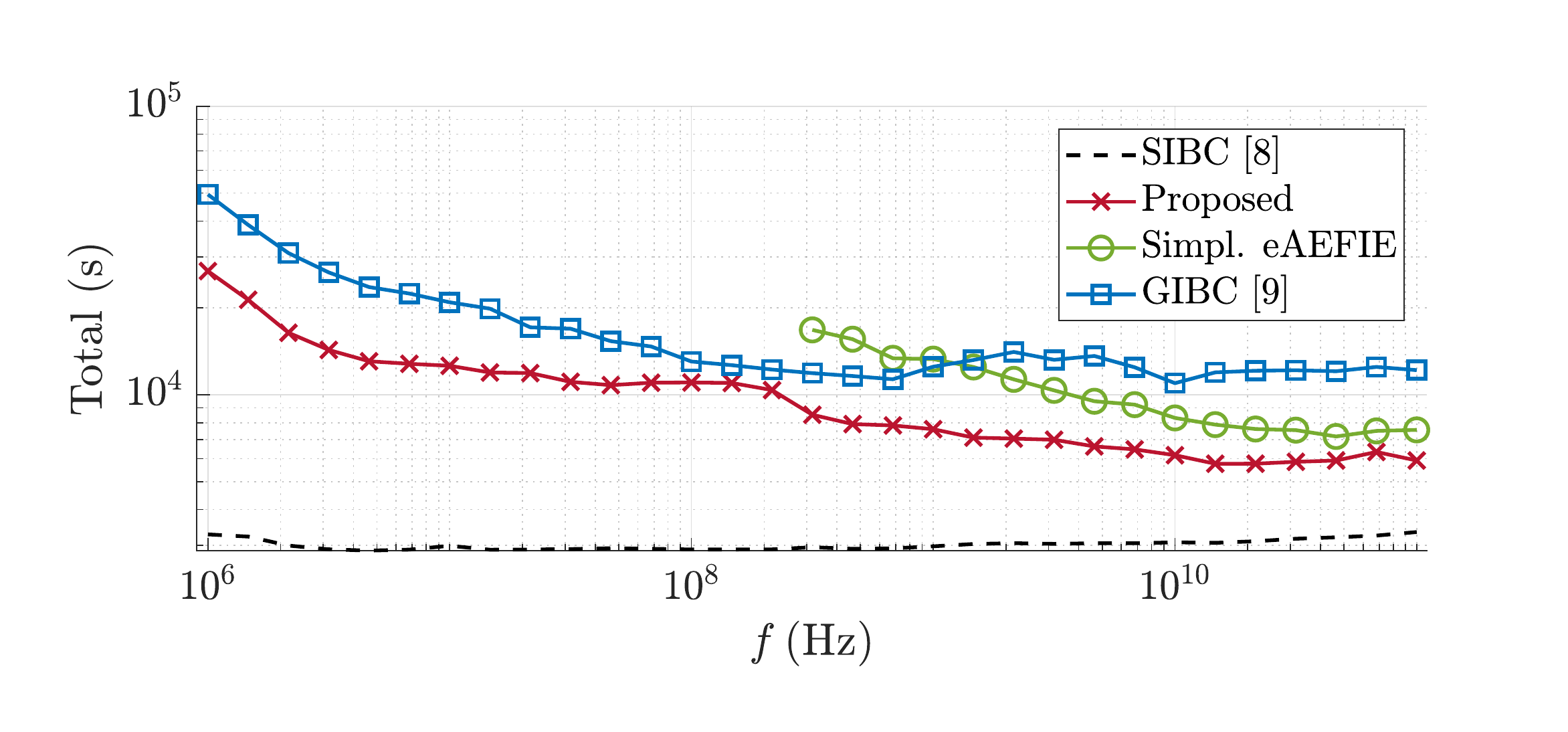}\label{fig:interposer_t_f}
	}
	\caption{Performance comparison for the interconnect network in \secref{sec:results:interposer}: (a) number of GMRES iterations required, (b) time taken in the interior problem, and (c) total time.}\label{fig:interposer_performance}
\end{figure}

\begin{figure}[t]
	\centering
	\includegraphics[width=\linewidth,trim={0 8mm 0 10mm},clip=true]{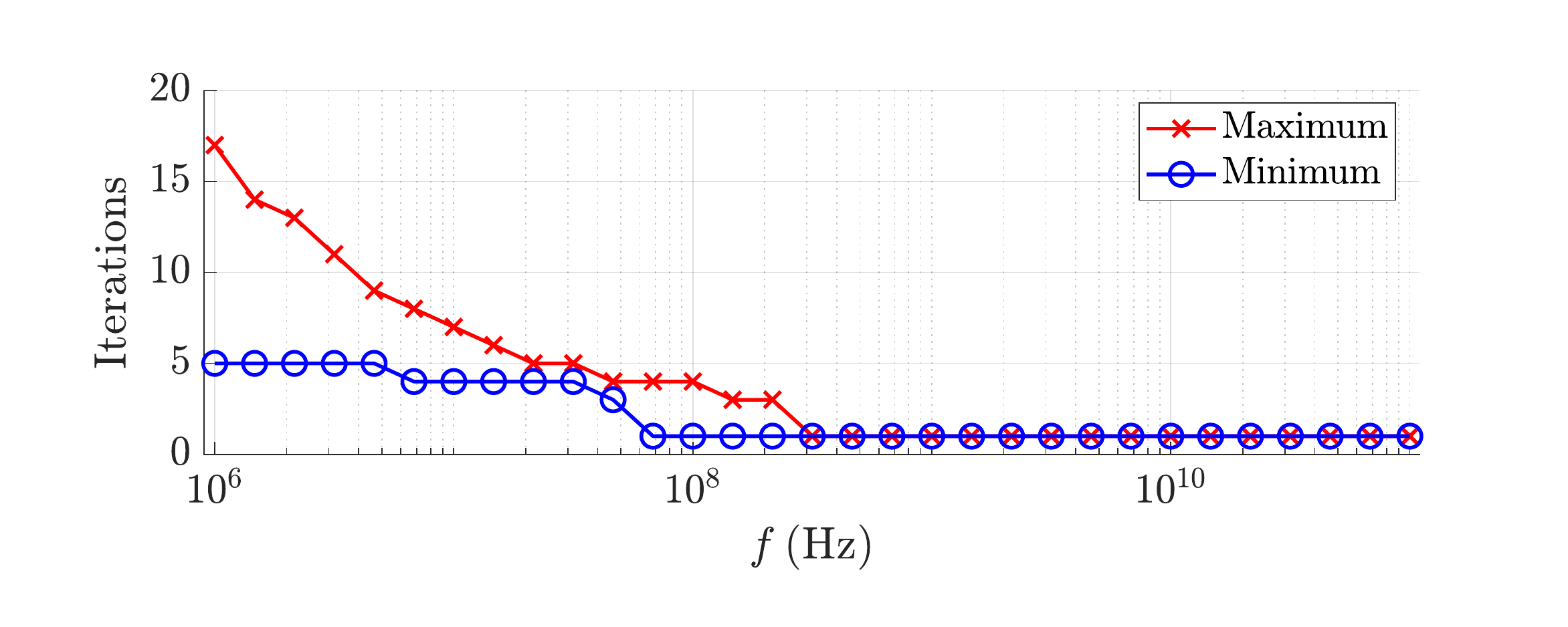}
	\caption{Maximum and minimum number of GMRES iterations required for the interior problem in the proposed method, for the interconnect network in \secref{sec:results:interposer}.}\label{fig:interposer_N_iter_inner}
\end{figure}

The performance of each method follows similar trends to the test case in \secref{sec:results:pckg}.
The proposed method remains well conditioned over most of the frequency range, while the simplified eAEFIE converges only at high frequencies, as shown in \figref{fig:interposer_N_iter}.
Although the number of iterations required by the proposed method increases at very low frequencies, its convergence is still significantly faster than the simplified eAEFIE.
\figref{fig:interposer_t_InnerModelGen} shows the cost of the interior problem per frequency.
In this case, the objects have similar sizes and are not very large compared to the full structure.
Still, the proposed method is significantly faster than direct GIBC by a factor, on average, of $10\times$.
The ability of the proposed method to take advantage of skin effect-induced matrix sparsity is clearly seen at high frequencies.
The overall improvement in total simulation time of the proposed method is evident in \figref{fig:interposer_t_f}.
Furthermore, \figref{fig:interposer_N_iter_inner} shows the maximum and minimum number of nested iterations required in solving~\eqref{eq:nest2}.
The maximum occurs for the ground plane, which is the largest object.
\figref{fig:interposer_N_iter_inner} confirms that the nested problem is efficiently solved with the near-region preconditioner.

\subsection{Discussion}\label{sec:results:discussion}
The reported results convey the following key observations:
\begin{itemize}
	\item The system resulting from the proposed method shows good convergence for a broad range of problems, without having to employ expensive dual basis functions.
	However, it should be expected that the introduction of dual basis functions may further improve convergence in the iterative solutions of both~\eqref{eq:sys} and~\eqref{eq:nest2}.
	Therefore, the proposed method with dual basis functions should still be competitive with the original eAEFIE.
	\item The proposed method allows accurate modeling of the interior problem without the explicit assembly of an impedance operator, unlike the original GIBC formulation.
	The proposed method eliminates the need to factorize dense matrices, and provides a significant improvement in CPU time over the original GIBC, especially when objects are large or not highly conductive.
	\item The proposed multiple-grid AIM accurately models dielectrics, lossy dielectrics, and conductors, over a broad range of frequency.
	Numerical results support the claim in \secref{sec:objAIM} that the AIM can be applied to lossy media without noticeable loss in accuracy.
	\item The near-region preconditioner~\eqref{eq:nest2pc} ensures that the nested system~\eqref{eq:nest2} is solved in a small number of iterations not exceeding $15$--$20$, while taking advantage of matrix sparsity due to the skin effect, when present.
\end{itemize}
The proposed technique successfully addresses the gaps in existing techniques laid out in \secref{sec:existing}, while taking advantage of their respective strengths.



\section{Conclusion}\label{sec:concl}
A fast surface integral equation technique based on an iteratively-applied surface impedance operator is proposed, for the unified modeling of penetrable media, from perfect dielectrics to good conductors.
The explicit assembly of the impedance operator is avoided, and its product with a vector is accelerated with a multiple-grid adaptive integral method (AIM).
The proposed method does not require assembling and factorizing dense matrices, and leads to a well conditioned system without the need for dual basis functions, which are expensive and increase code complexity.
The accuracy of the AIM for lossy materials is analyzed mathematically, and its applicability to conductive media is established over a broad frequency range.
The proposed technique is applied to a variety of realistic numerical examples, including a metasurface array, an integrated circuit package, and an on-chip interconnect network, and is shown to be significantly more efficient than existing methods.

\section*{Acknowledgment}

The authors would like to thank Advanced Micro Devices, and the anonymous reviewers for their thoughtful and constructive feedback.

\ifCLASSOPTIONcaptionsoff
  \newpage
\fi



\bibliographystyle{ieeetr}
\bibliography{IEEEabrv,./bibliography}

\begin{IEEEbiography}[{\includegraphics[width=1in,height=1.25in,clip,keepaspectratio]{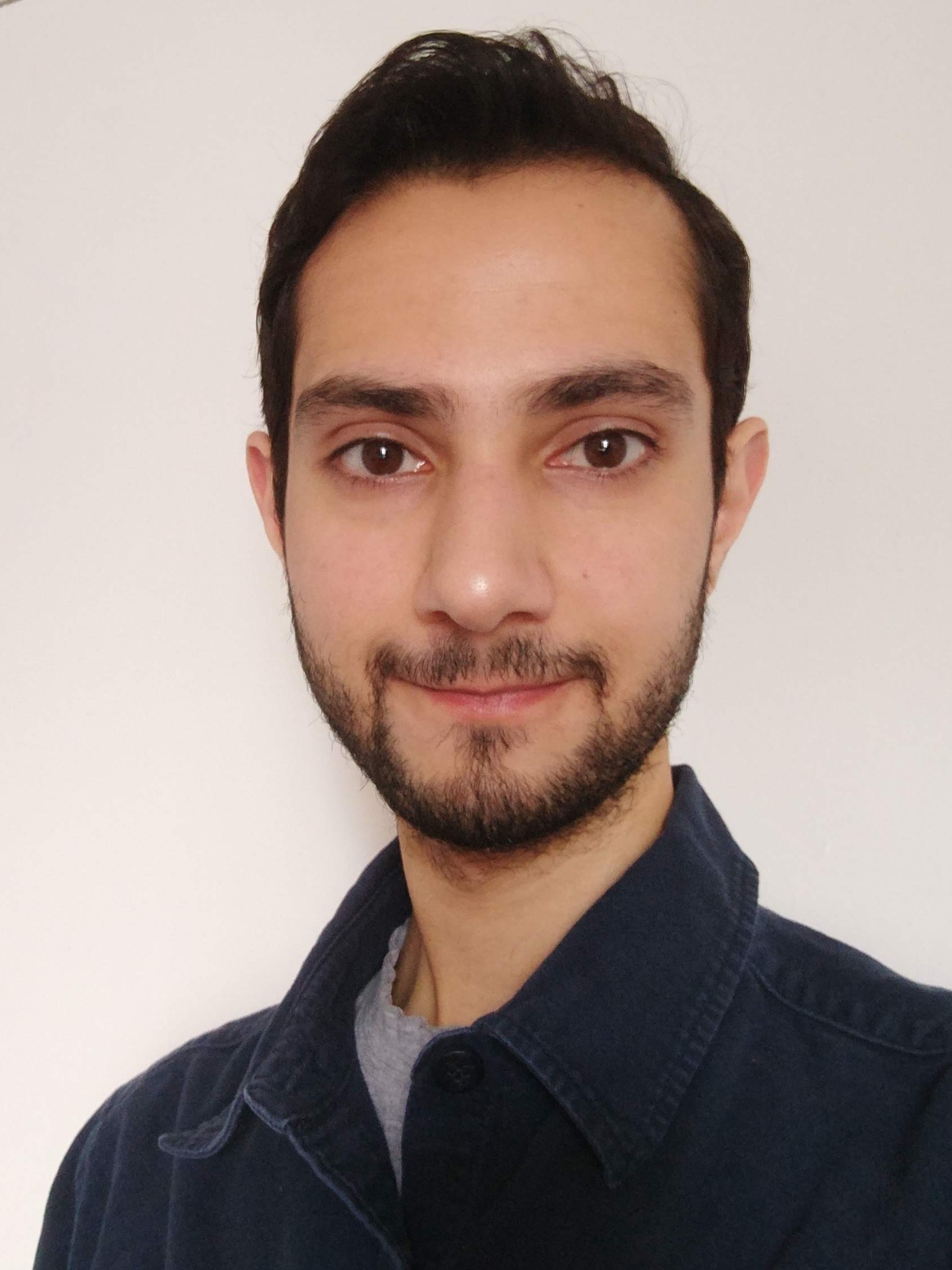}}]{Shashwat Sharma} (S'18)
received the B.A.Sc. degree in engineering physics and the M.A.Sc. degree in electrical engineering from the University of Toronto (U of T), Canada, in 2014 and 2016, respectively. From September 2016 to March 2017 he worked as a research intern in computational science at Autodesk, Toronto. Since 2017 he has been working towards the Ph.D. degreee in electrical engineering at U of T. His research focuses on computational electromagnetics, with an emphasis on fast and robust integral equation methods for multiscale electromagnetic modeling. His interests include all aspects of mathematical modeling and scientific computing applied to electromagnetics.

Mr.~Sharma placed second at the CNC/USNC-URSI Student Paper Competition of the 2020 IEEE International Symposium on Antennas and Propagation and North American Radio Science Meeting (AP-S/URSI), and was a finalist for the Best Student Paper Award at the IEEE Conference on Electrical
Performance of Electronic Packaging and Systems (2018). He also received an honorable mention for his contributions to the AP-S/URSI symposia in 2019 and 2020. He is currently serving as the vice chair for the U of T student chapter of the IEEE Antennas and Propagation Society.
\end{IEEEbiography}

\begin{IEEEbiography}[{\includegraphics[width=1in,height=1.25in,clip,keepaspectratio]{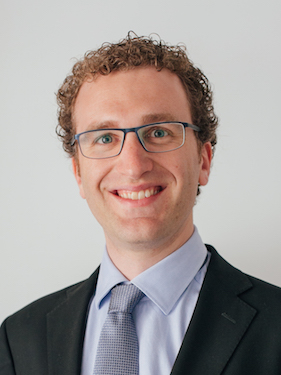}}]{Piero
Triverio} (S'06 -- M'09 -- SM'16)
received the Ph.D. degree in Electronic Engineering from Politecnico di
Torino, Italy, in 2009. He is an Associate Professor in The Edward S.
Rogers Sr. Department of Electrical \& Computer Engineering (ECE) at the
University of Toronto, and in the Institute of Biomaterials and
Biomedical Engineering (IBBME). He holds the Canada Research Chair in
Computational Electromagnetics. His research interests include signal
integrity, computational electromagnetism, model order reduction, and
computational fluid dynamics applied to cardiovascular diseases.

Prof.~Triverio received the Best Paper Award of the IEEE Transactions on
Advanced Packaging (2007), the EuMIC Young Engineer Prize (2010),  the
Connaught New Researcher Award (2013), and the Ontario Early Researcher
Award (2016). From 2013 to 2018, Triverio held the Canada Research Chair
in Modeling of Electrical Interconnects. Triverio and his students were
awarded the Best Paper Award of the IEEE Conference on Electrical
Performance of Electronic Packaging and Systems (2008, 2017), and
several Best Student Paper Awards at international symposia. He serves
as an Associate Editor for the IEEE TRANSACTIONS ON COMPONENTS,
PACKAGING AND MANUFACTURING TECHNOLOGY. He is a member of the Technical
Program Committee of the IEEE Workshop on Signal and Power Integrity,
and of the IEEE Conference on Electrical Performance of Electronic
Packaging and Systems.
\end{IEEEbiography}


\end{document}